\documentclass[sigplan,twocolumn]{acmart}
\acmSubmissionID{1113}

\makeatletter
\def\@ACM@checkaffil{}
\makeatother

\setcopyright{none}
\renewcommand\footnotetextcopyrightpermission[1]{}
\usepackage[english]{babel}
\usepackage{amsmath}
\usepackage{graphicx}
\usepackage{subcaption}
\usepackage{multirow}
\usepackage{array}
\usepackage{tcolorbox}
\usepackage{adjustbox}
\usepackage{makecell}
\usepackage{xcolor}
\usepackage{enumitem}
\usepackage{cleveref}
\usepackage{algorithm}
\usepackage{algorithmic}
\usepackage{booktabs}
\usepackage{xurl}
\usepackage{tikz}
\usepackage{eso-pic}
\usepackage{listings}
\usepackage{balance}
\AddToShipoutPicture*{
  \AtPageUpperLeft{
    \hspace{19.5cm}
    \raisebox{-1.9cm}{\includegraphics[width=0.07\textwidth]{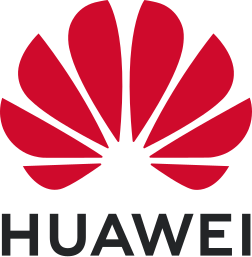}}
  }
}

\tcbset{
  insight/.style={
    colback=white,
    colframe=black,
    boxrule=0.4pt,
    left=1mm, right=1mm, top=0.5mm, bottom=0.5mm,
    boxsep=0.5mm
  }
}
\setlist[itemize]{leftmargin=*, label=\small$\bullet$}
\setlist[enumerate]{leftmargin=*}

\acmConference[]{}{}{}
\acmBooktitle{}
\acmYear{}
\acmDOI{}
\acmISBN{}

\begin{document}

\title{ElasticMoE: An Efficient Auto Scaling Method for Mixture-of-Experts Models}

\author{Gursimran Singh}
\authornote{Equal Contribution.}
\affiliation{\institution{Huawei Technologies Canada}}

\author{Timothy Yu}
\authornotemark[1]
\affiliation{\institution{Huawei Technologies Canada}}

\author{Haley Li}
\affiliation{\institution{Huawei Technologies Canada}}

\author{Cheng Chen}
\affiliation{\institution{Huawei Technologies Canada}}

\author{Hanieh Sadri}
\affiliation{\institution{Huawei Technologies Canada}}

\author{Qintao Zhang}
\affiliation{\institution{Huawei Technologies China}}

\author{Yu Zhang}
\affiliation{\institution{Huawei Technologies China}}

\author{Ying Xiong}
\affiliation{\institution{Huawei Technologies Canada}}

\author{Yong Zhang}
\authornote{Corresponding authors (<yong.zhang3, zhenan.fan1>@huawei.com).}
\affiliation{\institution{Huawei Technologies Canada}}

\author{Zhenan Fan}
\authornotemark[2]
\affiliation{\institution{Huawei Technologies Canada}}

\renewcommand{\shortauthors}{Singh et al.}

\begin{abstract}
Mixture-of-Experts (MoE) models promise efficient scaling of large language models (LLMs) by activating only a small subset of experts per token, but their parallelized inference pipelines make elastic serving challenging. Existing strategies fall short: horizontal scaling provisions entire replicas of the current configuration, often tens to hundreds of accelerators, leading to coarse granularity, long provisioning delays, and costly overprovisioning; vertical scaling offers finer adjustments but typically requires instance restarts, incurring downtime. These limitations make current approaches ill-suited for the bursty, short-lived traffic patterns common in cloud deployments.

We present ElasticMoE, an elastic scaling framework for MoE LLMs that achieves fine-grained, low latency, and zero-downtime scaling. ElasticMoE decouples inference execution from memory operations, enabling scaling steps to proceed concurrently with serving. An HBM Management Module (HMM) reuses weights and KV caches via zero-copy remapping, while high-bandwidth peer-to-peer transfers bring newly added accelerators online without interrupting service. A virtual memory–based expert redistribution mechanism migrates MoE experts without costly buffer reallocations, reducing peak memory usage during expert parallelism reconfiguration.

Our evaluation on Ascend NPUs with three popular MoE LLMs shows that ElasticMoE achieves up to $\approx$9X lower scale-up latency, up to $\approx$2X better throughput during scaling, and results in significant improvement in SLO attainment compared to baselines. By enabling fine-grained, concurrent scaling with minimal disruption, ElasticMoE advances the practicality of deploying massive MoE LLMs in dynamic cloud environments.

\end{abstract}

\maketitle
\pagestyle{plain}

\section{Introduction}
Mixture-of-Experts (MoE) models, such as DeepSeek V3~\cite{liu2024deepseek} and Qwen variants \cite{qwen2.5, qwen3}, have emerged as a compelling architecture for growing parameter counts without proportional compute cost. By activating only a subset of experts per token, MoE models significantly reduce per-inference computation while preserving the expressive power of massive parameter counts. This efficiency advantage has fueled their adoption in enterprise automation, code generation, and conversational AI, where state-of-the-art performance must coexist with practical serving costs.

\begin{figure}
  \centering
 \begin{subfigure}[t]{0.49\linewidth}
  \includegraphics[width=\linewidth]{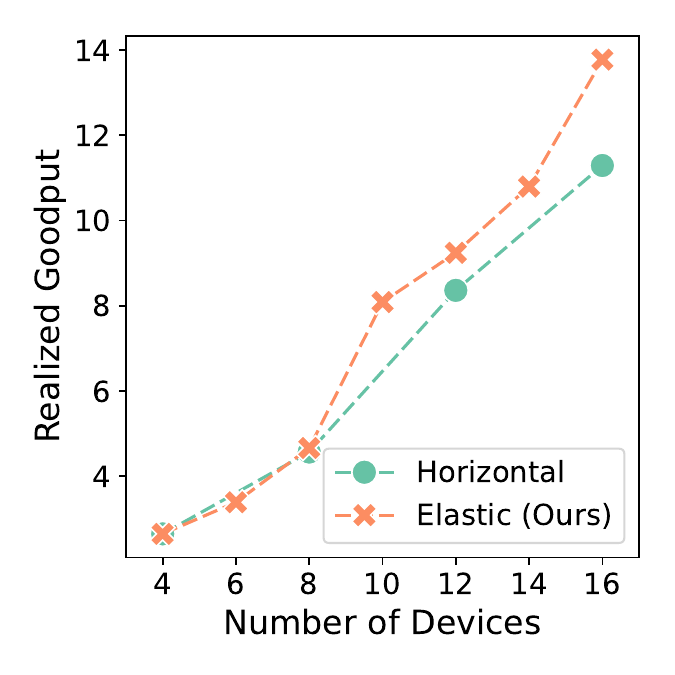}
  \caption{Achievable throughput given the number of devices.}
  \label{fig:motivation_rps:1}
  \end{subfigure}
  \hfill
 \begin{subfigure}[t]{0.49\linewidth}
  \includegraphics[width=\linewidth]{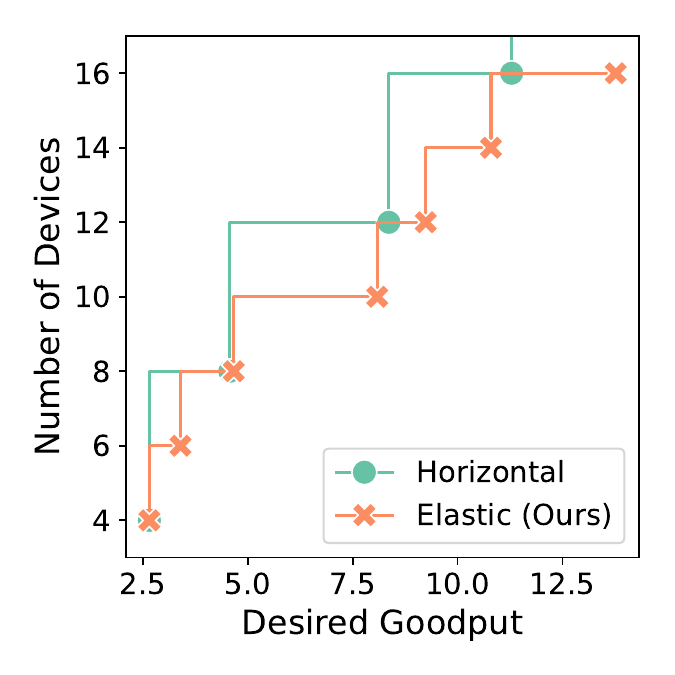}
  \caption{Hardware required to achieve a desired goodput.}
  \label{fig:motivation_rps:2}
  \end{subfigure}
  \caption{ElasticMoE (proposed) achieves better goodput (in terms of RPS) (a) and requires less hardware (b) due to more granular and flexible scaling for MoE models.}
  \label{fig:motivation_rps}
\end{figure}

However, serving large models in cloud environments poses unique challenges. Cloud workloads often exhibit highly variable and bursty request patterns ~\cite{khare2025superserve}, which means that serving systems must balance two competing objectives: sustaining high Service Level Objective (SLO) attainment and minimizing infrastructure cost. Achieving these goals requires the ability to adapt rapidly to fluctuating demand by scaling resources up during traffic spikes and scaling down during idle periods to avoid waste.

Current autoscaling strategies are dominated by horizontal scaling (scale-out/scale-in), which provisions or decommissions independent inference pipelines, each representing a full model instance spread across many accelerators. While simple to orchestrate, this approach has several drawbacks. First, it is coarse-grained. Scaling requires adding or removing a fixed quantum of accelerators. For example, even a minimal DeepSeek V3 inference instance may span 32 accelerators \cite{deepseekai2025deepseekv3technicalreport}, making finer adjustments infeasible. As a result, modest increases in traffic often force substantial overprovisioning, driving up infrastructure cost as shown in Fig.~\ref{fig:motivation_rps:2}. Second, instance startup sequences, such as container instantiation, model weight loading, communication setup, and KV-cache allocation can take tens of minutes for large models, which introduces significant scaling latency. This inertia prevents timely responses to short-lived traffic bursts, degrading SLO attainment during those periods. Finally, independently scaled instances in MoE models replicate expert-layer parameters, which dominate the model size~\cite{lepikhin2021gshard, deepseekai2025deepseekv3technicalreport, kimiteam2025kimik2openagentic}. This redundancy wastes memory that could otherwise be used for activations or KV-cache, lowering throughput and SLO efficiency as shown in Fig.~\ref{fig:motivation_rps:1}. 

\begin{figure}
  \centering
  \includegraphics[width=\linewidth]{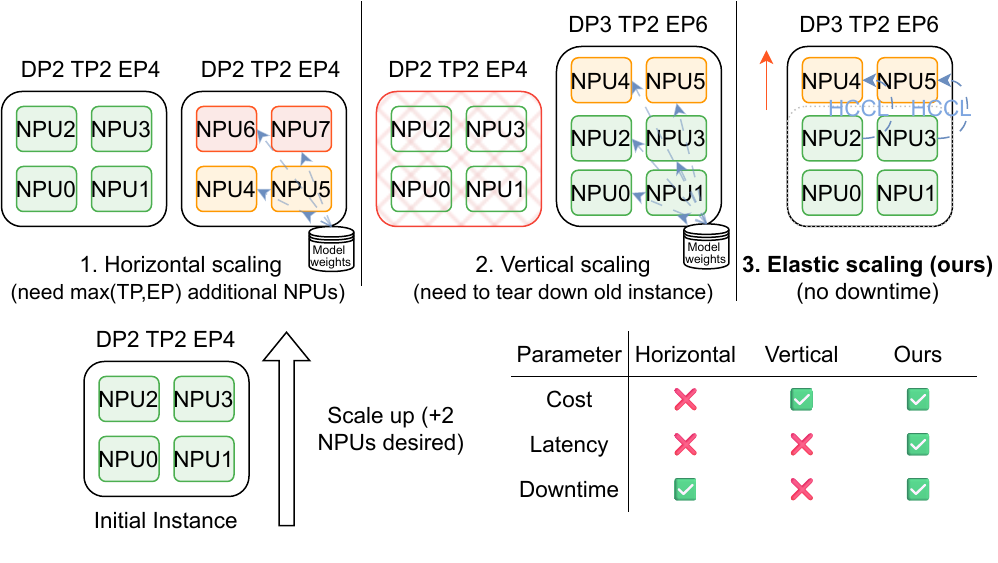}
  \caption{Comparison of scaling methods. Horizontal scaling adds a full replica, requiring coarse-grained capacity increases. Vertical scaling resizes an instance but requires cold restart incurring downtime. ElasticMoE scales in place, avoiding both inefficiency and downtime.}
  \label{fig:elastic_tf}
\end{figure}

In contrast, vertical scaling (scale-up/scale-down) adjusts the resource footprint of an existing inference instance by adding or removing only a few accelerators—for example, scaling from 32 NPUs to 34 NPUs. While more fine-grained, naïve approaches lack support for live reconfiguration. They typically require tearing down the current instance and restarting it with the new configuration, incurring downtime and cold-start latency. Other methods attempt to launch the scaled-up instance concurrently on the same accelerators, but this creates peak memory spikes as old and new instances temporarily coexist. Both approaches undermine the benefits of elastic scaling in cloud environments, where downtime or memory pressure during reconfiguration can further exacerbate SLO violations. 

These trade-offs are illustrated in Fig.~\ref{fig:elastic_tf}, which shows horizontal scaling’s coarse granularity and high cost with vertical scaling’s downtime and high latency. To cope with these limitations, production systems often adopt conservative fallback policies such as static overprovisioning or aggressive cooldown timers. These reduce the frequency of autoscaling events but at the cost of resource inefficiency and sluggish adaptation. Such rigidity is particularly problematic under bursty or short-lived traffic patterns, where effective scaling must be fast, precise, and minimally disruptive.

\begin{figure}[h]
  \centering
  \includegraphics[width=0.9\columnwidth]{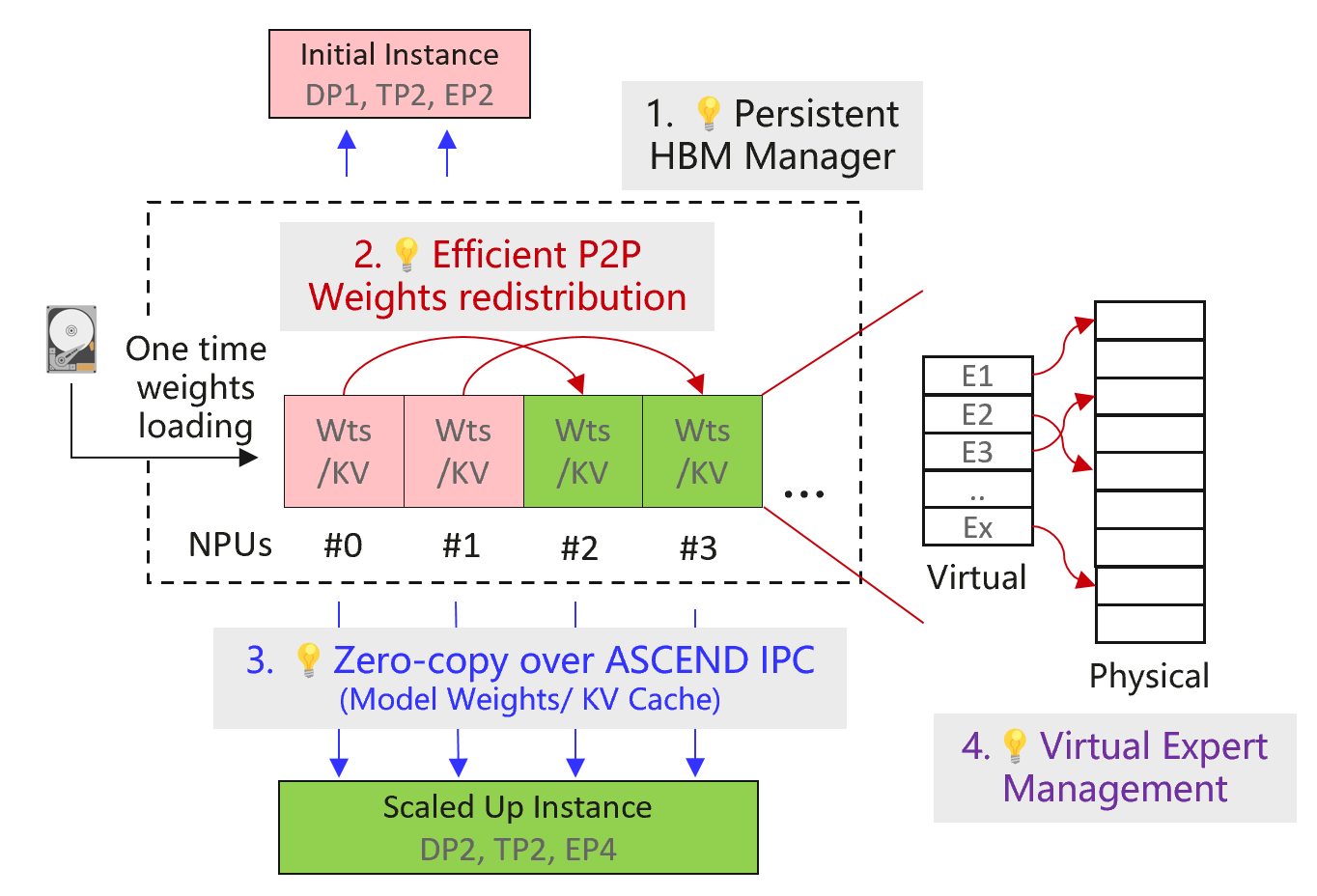}
  \caption{Key innovations of ElasticMoE: (i) decoupled HBM management from inference, (ii) zero-copy reuse of weights and KV-caches along with high-speed P2P transfers for reconfiguration, and (iii) virtual expert managemnt.}
  \label{fig:key-innovations}
\end{figure}

\paragraph{Our approach:}
This paper introduces ElasticMoE, a system for elastically serving large-scale MoE models that enables low-latency, fine-grained, and zero-downtime vertical scaling in production. ElasticMoE builds on three key ideas, summarized in Fig.~\ref{fig:key-innovations}.

\textit{First}, ElasticMoE decouples inference execution from memory management tasks such as model weight loading and KV-cache setup. At its core is a memory manager for the high bandwidth memory (HBM) called the HBM Management Module (HMM), which manages model weights and KV-caches independently of inference instances. Instances do not allocate these resources directly but instead receive the corresponding ``pointers'' in HBM via Ascend’s IPC mechanism \cite{ascend2025ipc}. This decoupling allows the HMM to reconfigure asynchronously during scaling, enabling the active instance to continue serving requests without interruption.

\textit{Second}, ElasticMoE performs scaling in place while minimizing expensive weight movements and memory reallocations. Specifically, it adjusts data parallel (DP) and expert parallel (EP) degrees while keeping tensor parallel (TP) fixed. This design reduces inter-accelerator weight transfers and allows KV-caches on existing devices to be reused, ensuring uninterrupted inference. Reconfiguration is further accelerated by high-bandwidth peer-to-peer (P2P) transfers using Ascend’s HCCL \cite{ascend2025hccl}, resulting in low-latency scale-up.

\textit{Finally}, ElasticMoE employs a virtual memory and page-based expert weight management mechanism that supports dynamic MoE expert reconfiguration. By treating expert weights as a contiguous logical tensor mapped to underlying pages, experts can be remapped in place without large buffer reallocations or full weight copies. This approach lowers scaling latency, peak memory pressure, and preserves throughput stability during scaling.

\paragraph{Contributions.} We make the following contributions:

\begin{enumerate}
\item \textbf{Elastic scaling framework for MoE models.}  
We design \textit{ElasticMoE}, a novel system that incrementally resizes inference instances by adding or removing NPUs, enabling fine-grained vertical scaling beyond the coarse granularity of horizontal approaches. 

\item \textbf{Novel mechanisms for efficient scaling.}
ElasticMoE introduces several novel techniques: (i) decoupled memory and inference management, (ii) low-overhead reconfiguration using zero-copy sharing and high-bandwidth P2P transfers, and (iii) virtual expert management for efficient redistribution of MoE experts. Together, these mechanisms minimize scale-up latency, avoid downtime, and reduce peak memory overhead.  

\item \textbf{Prototype and evaluation.}  
We implement \textit{ElasticMoE} on a Huawei CloudMatrix384 supernode~\cite{zuo2025serving} and evaluate it on three large MoE LLMs across two workload settings. Results show substantial improvements in scaling latency ($\approx$9X better), memory usage, and throughput stability ( $\approx$2X better) compared to horizontal and vertical scaling baselines, while consistently sustaining target SLOs.  
\end{enumerate}

\section{Background}

\subsection{Distributed Inference for MoE Models} \label{moe-background}

Mixture-of-Experts (MoE) models extend transformers with sparsely activated expert layers. Each layer contains a shared attention module followed by many experts, of which only a few are selected per token via a gating function (e.g., 8 of 256 in DeepSeek V3~\cite{deepseekai2025deepseekv3technicalreport}). This conditional computation enables massive capacity at lower per-token compute cost.

Due to large parameter counts, KV caches, and activations, MoE inference typically spans multiple accelerators. Efficient deployment combines three forms of parallelism. \textit{Data parallelism (DP)} replicates attention and feed-forward modules to process independent batches concurrently. Within each replica, \textit{Tensor parallelism (TP)} shards large matrix operations (typically attention operations) across devices to reduce memory load. \textit{Expert parallelism (EP)} distributes experts across devices, with tokens dynamically routed to their assigned experts. A common configuration sets $\text{EP} = \text{TP} \times \text{DP}$, ensuring one expert per device, while other strategies replicate or unevenly distribute experts to balance load and mitigate stragglers~\cite{xiao2025xdeepserve, deepseekai2025deepseekv3technicalreport, wu2024lazarusresilientelastictraining}. These choices directly impact scalability, throughput, and cost efficiency.

\subsection{Autoscaling in Cloud Environments} \label{autoscale-background}

Autoscaling enables LLM serving systems to adapt resources to fluctuating demand while balancing cost and latency. Real-world traffic is highly bursty, with request rates surging by more than 10$\times$ within minutes in production deployments~\cite{yu2025lambda}. Without rapid scale-up, requests quickly overwhelm serving capacity and violate service-level objectives (SLOs). Conversely, when demand falls, systems must scale down promptly to avoid wasted resources.

The dominant approach is \textit{horizontal scaling}, which launches or removes full serving instances on independent nodes. For example, a minimal configuration of DP2-TP2-EP4 across 4 accelerators must be scaled out by adding an entirely new DP2–TP2–EP4 replica, doubling resource use regardless of actual demand. While this strategy is easy to integrate with orchestrators such as Kubernetes, Ray Serve~\cite{moritz2018ray}, and AWS SageMaker, it is coarse-grained and slow: each instance requires container startup and weight loading. For MoE models, it is further inefficient (Fig.~\ref{fig:model_memory} and Fig.~\ref{fig:motivation_rps:1}) since experts remain confined within isolated instances.

An alternative is \textit{vertical scaling}, which enlarges the resource footprint of a single instance, for example, reconfiguring from DP2–TP2–EP4 on 4 accelerators to DP3–TP2–EP6 on 6 accelerators. This offers finer granularity and greater flexibility in expert distribution, but typically requires tearing down and reinitializing the instance, incurring cold-start latency and significant downtime that can be detrimental for SLOs, defeating the very purpose of scaling.

Both strategies therefore trade off latency, granularity, flexibility, and overhead. Overcoming these limitations is essential for efficient autoscaling of MoE inference, as we discuss in the next section.

\section{Opportunities and Key Insights}  \label{sec:limitations}

LLM serving systems deployed in production must elastically scale to handle dynamic and often unpredictable workloads while maintaining strict performance guarantees. However, current autoscaling strategies—both horizontal and vertical—fall short in several key areas. We highlight four core limitations that motivate the need for a more responsive and resource-efficient solution.

\begin{figure}[h]
  \centering
  \begin{subfigure}[t]{0.59\linewidth}
    \includegraphics[width=\linewidth]{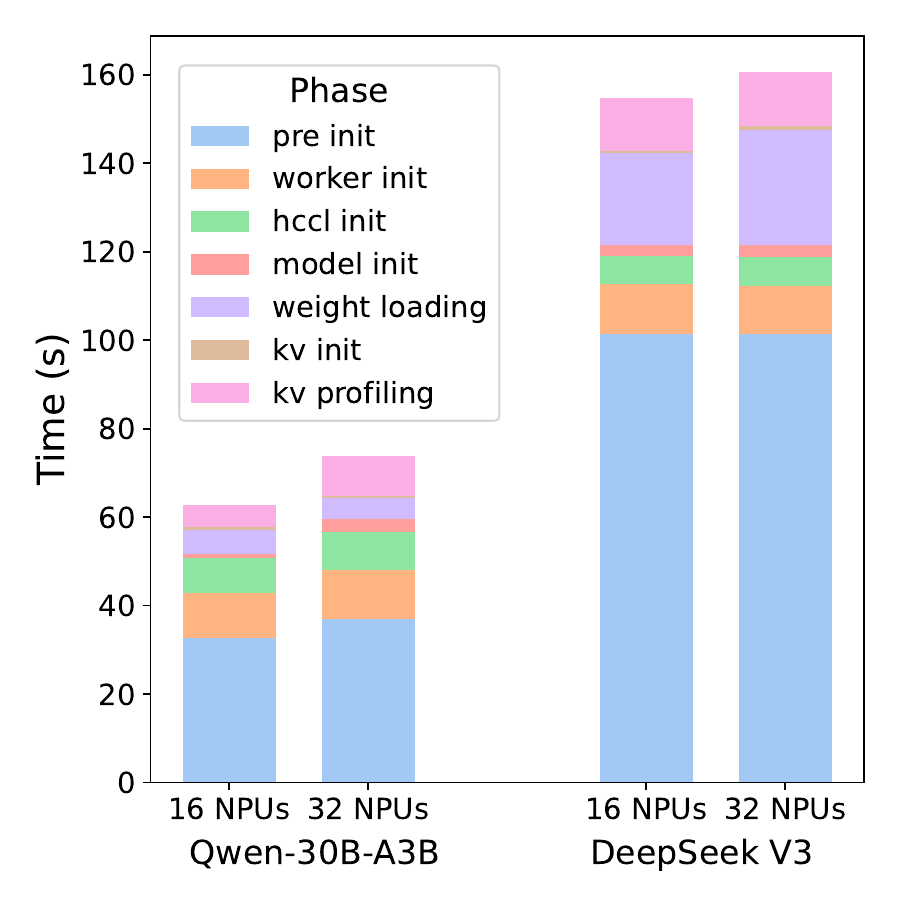}
    \caption{Instance initialization latency.} 
    \label{fig:init_breakdown}
  \end{subfigure}
  \hfill
  \begin{subfigure}[t]{0.39\linewidth}
    \includegraphics[width=\linewidth]{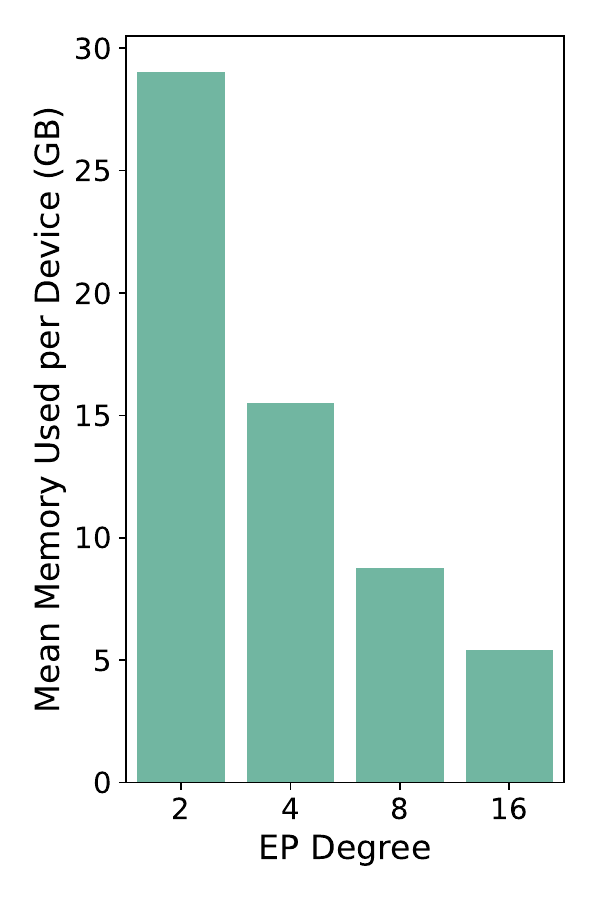}
    \caption{Model memory usage.}
    \label{fig:model_memory}
  \end{subfigure}
  \caption{Boot-up and memory analysis: (a) instance initialization latency breakdown, and (b) per-device memory consumption for model weights across EP degrees.}
  \label{fig:p2p-copy}
\end{figure}

\paragraph{L1: High scaling latency.}
Existing systems lack the responsiveness needed to handle sudden, short-lived spikes in request load—a frequent pattern in production deployments~\cite{yu2025lambda}. The root cause is cold starting the new instance that undergo long boot-up sequences, including weight loading, memory allocation, and communication initialization, which together introduce scaling latencies of tens of seconds to minutes (Fig.~\ref{fig:init_breakdown}). This high provisioning inertia causes requests during traffic spikes to suffer SLO violations due to inadequate resources, directly harming end-user experience.

\begin{tcolorbox}[insight]
\textbf{Insight 1:} 
\textit{Booting up new instances adds substantial delays that worsens with model size and accelerator count; scaling must therefore avoid naive cold-starts to remain responsive.}
\end{tcolorbox}

\paragraph{L2: High downtime during scaling.}
Vertical scaling often requires tearing down the current instance before bringing up a new one, causing unavoidable service interruptions. During this transition, in-flight and incoming requests are dropped or delayed, leading to immediate SLO violations. Meanwhile, new requests continue to queue, creating a backlog that amplifies latency even after recovery and may require additional overprovisioning to drain. These cascading effects make traditional scaling unsuitable for environments with strict latency guarantees and continuous traffic.

\begin{tcolorbox}[insight]
\textbf{Insight 2:} 
\textit{Downtime during scaling is unacceptable for SLO compliance; scaling must therefore proceed concurrently while the active instance continues serving requests.}
\end{tcolorbox}

\paragraph{L3: Coarse scaling granularity.}
Large-scale MoE inference relies on extensive model parallelism, for example, a single DeepSeek V3 instance spans 32–320 accelerators across multiple nodes~\cite{deepseekai2025deepseekv3technicalreport}. Horizontal autoscaling must launch entire replicas of this size, so even a modest increase in demand requires allocating a minimum of 32–320 accelerators. This coarse scaling granularity prevents incremental adjustments: small spikes force massive overprovisioning, lowering resource efficiency, and inflating cost.

\begin{tcolorbox}[insight]
\textbf{Insight 3:} 
\textit{Scaling must support fine-grained resource adjustment so that small load fluctuations can be handled without massive overprovisioning.}
\end{tcolorbox}

\paragraph{L4: Inefficient expert redistribution.}
In MoE models, expert layers dominate model size and are therefore distributed across accelerators using expert parallelism (EP)~\cite{lepikhin2021gshard}. Naïve horizontal scaling replicates these experts in every new instance, effectively limiting the EP degree within each instance. This replication inflates memory consumption (Fig.~\ref{fig:model_memory}) and leaves less capacity for KV cache and activations, leading to smaller batch sizes, and degraded throughput (Fig.~\ref{fig:motivation_rps:1}. Furthermore, because each scaled instance operates in isolation, experts cannot be coordinated across instances. Without a unified token routing mechanism, load balancing~\cite{wu2024lazarusresilientelastictraining, deepseekai2025deepseekv3technicalreport} across accelerators is impeded, further diminishing the efficiency gains of MoE sparsity.

\begin{tcolorbox}[insight]
\textbf{Insight 4:} 
\textit{Scaling solutions for MoE must allow flexible expert redistribution across accelerators to avoid parameter duplication and unlock efficient load balancing.}
\end{tcolorbox}

\paragraph{L5: High peak memory requirement.}
To prevent downtime, some vertical scaling approaches launch the new configuration alongside the existing one~\cite{stojkovic2025dynamollm}. However, this raises the peak memory requirement during scaling that manifests in either of two ways. If the same accelerators are reused, they must temporarily hold two copies of the model, reducing space for KV caches and activations and risking OOM failures. Alternatively, launching the new configuration on additional accelerators avoids memory pressure but doubles resource consumption during the transition, inflating cost. Both approaches are inefficient—the first undermines stability, while the second sacrifices cost-efficiency.

\begin{tcolorbox}[insight]
\textbf{Insight 5:} 
\textit{Additional resources and high peak memory during scaling are undesirable; effective solutions must minimize peak memory by avoiding redundant allocation of model weights and KV caches on shared accelerators.}
\end{tcolorbox}

\section{ElasticMoE: System and Design}
Following the above insights, we advocate for an \textit{elastic scaling framework}, \textbf{ElasticMoE}, that enables fine-grained, low-latency, and zero-downtime scaling of MoE models in production. To achieve these goals, the system must satisfy several requirements: 
(i) enable fast transitions between configurations with minimal latency and memory overhead to remain responsive under bursty workloads;  
(ii) without using additional resources, support concurrent scaling while the active configuration continues serving requests, ensuring uninterrupted availability; and
(iii) provide MoE-aware, fine-grained allocation and deallocation of resources, so that scaling preserves memory efficiency, runtime efficiency, and cost-effectiveness.  
Meeting these requirements introduces several intertwined \textbf{challenges}:  

First, reducing scaling latency requires \textit{avoiding or minimizing costly boot-up tasks} such as model initialization, weight loading, and KV-cache setup. Achieving this is non-trivial, especially as these tasks are specific to the target configuration and are typically performed sequentially.
Second, the system must continue serving requests while concurrently bringing up a new configuration.
Coordinating execution and reconfiguration \textit{without using additional accelerators or high peak memory} during scaling is particularly challenging when token routing or expert redistribution are involved. Finally, supporting fine-grained scaling requires major adjustments to the serving instance, like resharding model weights, adjusting parallelism degrees, etc. Doing this \textit{without restarting the entire instance} or disrupting ongoing inference service is a major challenge.

\subsection{Design Choices}
In this section, we present key design choices that allow ElasticMoE to scale efficiently and asynchronously, with minimal peak memory overhead, scaling latency, and MoE-aware adaptability.

\paragraph{Scaling via DP and EP Reconfiguration.}
In MoE models, the model architecture combines shared attention layers with sparsely activated experts. Attention modules are typically partitioned with TP, while DP and EP determine the number of attention replicas and the distribution of experts across devices (see Section~\ref{moe-background}). In ElasticMoE, TP is fixed during scaling so that attention modules and KV caches remain unchanged on accelerators already in use. This allows the new configuration to be brought up without disrupting the active instance, while preserving consistency in shared components. Because the common accelerators in both configurations retain the same layout, part of model weights (except expert layers) and KV caches can be directly reused instead of reallocated, avoiding peak memory pressure. Scaling is therefore achieved by adjusting only the DP and EP degrees, enabling incremental resizing of inference instances with low latency, avoiding disruption, and minimal overhead.

\paragraph{OS-level Asynchronous Process Management.}
To support concurrent inference and scaling, ElasticMoE uses a process-level isolation strategy. When a scale-up operation is triggered, a new inference process is launched with access to a superset of the original accelerators (i.e., both shared and newly added devices). This process performs initialization—including model loading on new devices and communication group setup—independently, while the original inference process continues to serve requests. Once the new instance is ready, traffic is rerouted, and the old process is terminated. This ``scale-while-serve'' model avoids service downtime and enables scaling to occur without blocking ongoing inference, even under bursty workloads.

\paragraph{Decoupled Memory and Execution Layers.}
ElasticMoE adopts a two-tier architecture that separates HBM management from inference execution. A dedicated memory management daemon is responsible for loading model weights and KV caches, maintaining them in device memory, and applying any reconfigurations. Execution processes, by contrast, never load weights or KV caches directly; instead, they obtain reference handles to the tensors managed by the memory daemon. This decoupling offers two main advantages. First, inference instances can be started or shut down without forcing reloads, since the memory state persists independently. 
Second, reconfiguration can be carried out efficiently and asynchronously: the daemon knows the current and the target configuration and can apply minimal changes to provision newly added devices, avoiding costly disk I/O and ensuring that active inference continues without disruption.

\paragraph{Fast Model Loading and KV Cache Initialization.}
As shown in Fig.~\ref{fig:init_breakdown}, naively loading model weights from disk during scaling is prohibitively expensive. ElasticMoE accelerates this process with two key techniques. First, devices shared between the old and new configurations reuse existing weight tensors and KV caches via zero-copy mechanisms, making scaling on these devices essentially free. Second, for newly added accelerators, ElasticMoE performs only the necessary peer-to-peer (P2P) transfers over high-bandwidth interconnects such as the Ascend Unified Bus or RDMA-capable links. The system explicitly plans scaling transitions to maximize zero-copy reuse and restrict weight transfers to the minimal required set of devices. Together, these optimizations eliminate redundant disk I/O, lower scale-up latency, and mitigate peak memory pressure during transitions.

\paragraph{Virtual Expert Management.}
Expert weights in MoE layers are stored as large contiguous tensors for kernel efficiency. During scaling, such as changing from \texttt{EP=4} to \texttt{EP=6}, some experts must be migrated to new devices while the remaining on existing devices are rebuilt from a subset of the original experts. Naïvely reshaping these tensors requires allocating new contiguous buffers and copying large portions of data in memory, which both increases latency and creates high peak memory demand. ElasticMoE addresses this with a virtual memory abstraction that enables $O(1)$ expert reshaping. Experts are stored as non-contiguous pages in physical memory but are mapped into contiguous regions in virtual memory, satisfying kernel requirements while avoiding bulk copies. This allows experts to be remapped dynamically during scaling with minimal latency and without incurring memory pressure.

\begin{figure*}[htbp]
  \centering
  \includegraphics[width=0.9\textwidth]{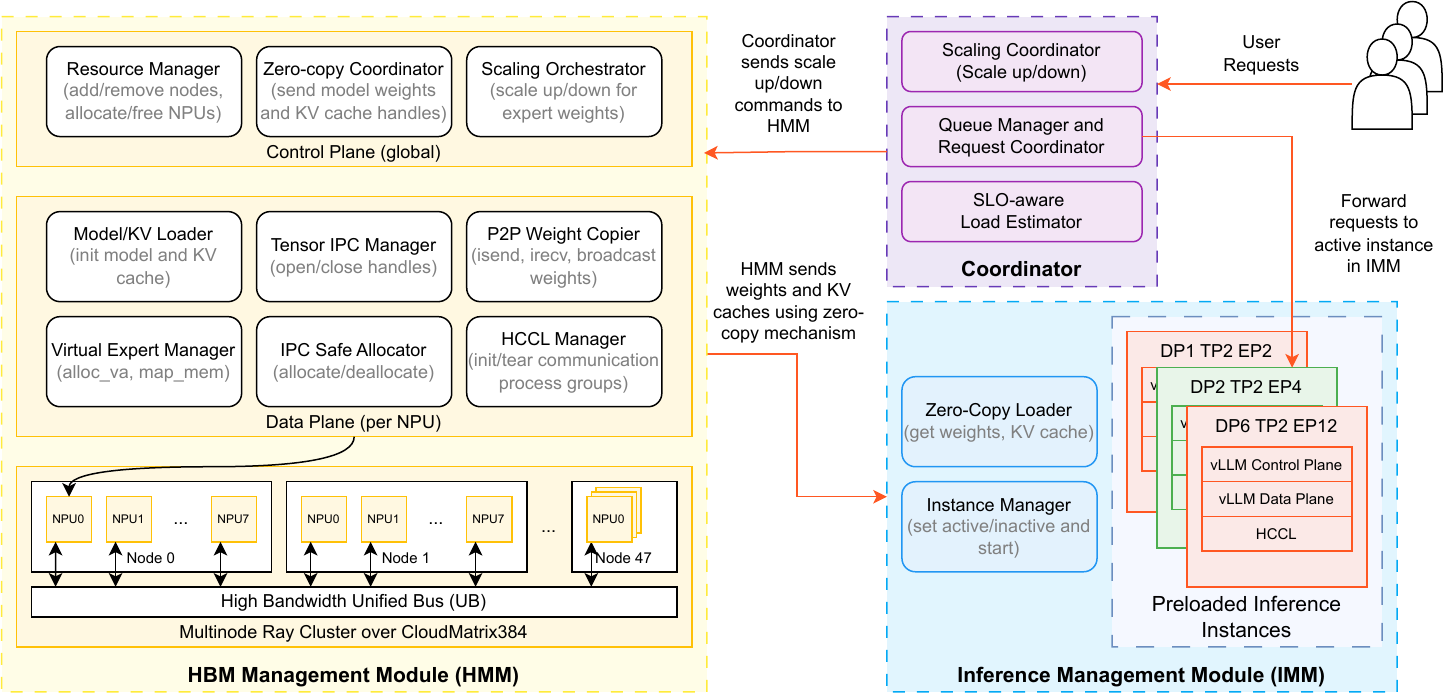}
  \caption{System architecture of ElasticMoE with three components: (i) persistent HMM for weight and KV-cache management, (ii) Coordinator for scheduling, SLO monitoring, and scaling, and (iii) IMM with transient preloaded but selectively activated instances. Zero-copy sharing and high-bandwidth transfers enable fine-grained, low-latency scaling without downtime.}
  \label{fig:architecture}
\end{figure*}

\subsection{System Overview}
Fig.~\ref{fig:architecture} presents an architectural overview of \textit{ElasticMoE}. The system is organized into three main components: the \textit{Coordinator}, the \textit{HBM Management Module} (HMM), and the \textit{Inference Management Module} (IMD).

The \textbf{Coordinator} serves as the entry point for user requests, routing queries to the active inference instance and orchestrating scaling decisions by coordinating the HMM and IMM. The \textbf{HMM} manages model weights and KV caches in device memory, separating costly initialization from inference execution. It loads weights once, maintains them persistently, and enables efficient scaling through zero-copy reuse and high-speed peer-to-peer transfers. The \textbf{IMM} runs inference, keeping multiple pre-initialized instances ready but activating only one at a time. It obtains model weights and KV caches for the active instance through zero-copy reference handles provided by the HMM. During scaling, the IMM prepares the target configuration while the current instance continues serving requests, ensuring uninterrupted service. In the following subsections, we describe these components in greater detail.

\subsection{Coordinator}
It acts as the entry point for user queries and manages the overall control flow of \textit{ElasticMoE}. It maintains an active request queue, monitors SLOs, and routes queries to the currently active inference instance. At initialization, the Coordinator prepares the runtime environment to ensure that scaling operations can be executed efficiently.

A key component of the Coordinator is the \textit{SLO-aware Load Estimator}, which continuously tracks SLO attainment rates. When the estimator detects persistent violations (e.g., SLO attainment falling below 90\%), it triggers a scale-up command; conversely, it can initiate scale-down to reduce overprovisioning during low-load periods. Once the target configuration is ready, the Coordinator seamlessly redirects traffic from the old instance to the new one, ensuring uninterrupted service and stable request latency.

\subsection{HBM Management Module (HMM)}

The HMM is the core of \textit{ElasticMoE}, responsible for managing HBM initialization state operations like model weights and KV caches in device memory and decoupling these expensive operations from inference execution. It loads weights and initializes KV caches only once, keeps them persistently available, and reuses them across inference instances in IMM. During inference, HMM, shares the loaded model weights and KV caches to active instance in IMM. During scaling, HMM computes a scaling plan that allows reuse of HBM state on shared accelerators and provisions new devices through high-speed peer-to-peer transfers, avoiding repeated disk I/O. This design minimizes redundant weight movement, reduces peak memory usage, and ensures that scaling transitions can be completed with minimal latency while inference continues uninterrupted.

HMM is organized into a \textit{global control plane} and a set of \textit{per-device workers}, both implemented on top of a distributed runtime (Ray). The control plane maintains cluster-wide state and coordinates scaling by tracking resources, distributing zero-copy references to active inference instance in IMM, and orchestrating when and how scaling actions occur. The per-device workers, each bound to a physical accelerator, carry out the actual data-plane operations: loading model weights and KV caches, managing inter-process memory handles using Ascend IPC, performing peer-to-peer weight transfers using HCCL, and remapping experts during EP reconfiguration. More details about these operations are explained in §~\ref{low-level-premitives}. Communication between the control plane and workers is exposed through Ray primitives, enabling fine-grained, asynchronous operations on each accelerator under centralized coordination.

\subsection{Inference Management Module (IMM)}

The IMM is responsible for executing model inference using the weights and KV caches provided by the HMM. Unlike traditional systems where each instance must independently load weights and initialize caches, ElasticMoE allows IMM processes to attach to existing memory through zero-copy reference handles. This eliminates redundant initialization, reduces startup latency, and ensures that active instances can immediately begin serving requests with minimal overhead.

IMM maintains multiple inference instances but activates only one at a time. Idle instances can be pre-initialized on CPUs for anticipated configurations, kept in a standby state, and switched in with minimal delay when scaling is triggered. This design allows ElasticMoE to prepare a new configuration while the current one continues serving queries, avoiding downtime. Once the new instance is fully ready, the IMM seamlessly transitions user traffic to it and retires the old configuration.

Internally, the IMM provides two key functions. First, the \textit{Zero-Copy Loader} retrieves model weights and KV caches from the HMM through reference handles, replacing the standard disk-based weight loader (\texttt{DiskLoader}) used in systems like vLLM. Second, the \textit{Instance Manager} tracks the lifecycle of inference instances, marking them active or inactive, maintaining an LRU cache of standby instances, and orchestrating smooth handoff during scale-up or scale-down.

\subsection{Low-level Primitives} \label{low-level-premitives}

ElasticMoE builds on a set of lightweight primitives that enable memory reuse, fast weight movement, and efficient expert redistribution across NPUs. These primitives are implemented in C++ within the HMM and IMM's data plane and exposed to Python via PyBind, forming the foundation for fine-grained, low-latency scaling. Below, we briefly describe the core APIs; more details are provided in Appendix~\ref{low-level-premitives-implementation-details}.

\paragraph{zero-copy.} Enables sharing of tensors across multiple processes using the Ascend IPC mechanism. Tensors are first allocated with an \texttt{IpcSafeAllocator}, which replaces PyTorch’s default \texttt{TorchCachingAllocator} and ensures allocations are IPC-compatible in device memory. The HMM exports handles via \texttt{export\_handle}, which the IMM can open through the \texttt{open\_tensor} primitive in the ZeroCopyLoader. This eliminates redundant copies and reduces peak memory usage during scaling.

\paragraph{p2p-copy.} Provides high-speed peer-to-peer transfers between devices using Ascend’s HCCL collective communication library. As a one-time setup, the HMM initializes an HCCL domain across all devices and nodes with \texttt{init\_\allowbreak process\_\allowbreak group}. During scaling, tensors are moved using HCCL directives such as \texttt{isend}, \texttt{irecv}, and \texttt{broadcast}. Transfers occur over the Ascend Unified Bus (or RDMA links), bypassing host memory and minimizing latency.

\paragraph{vpage-remap.} Introduces a virtual memory abstraction in which expert weights are split into physical pages but exposed contiguously in virtual address space. Pages allocated with the \texttt{MallocPhysical} are grouped into a virtual range using \texttt{alloc\_va}, and mapped into this space with \texttt{map\_mem}. This enables $O(1)$ expert reshaping with minimal latency and peak memory overhead during EP reconfiguration.

\section{Elastic Inference Lifecycle}

In this section, we describe the lifecycle of inference in \textit{ElasticMoE}. We begin with the \textit{initialization process}, where the system boots up and starts serving with the initial configuration. We then explain how the lifecycle evolves during a \textit{scale-up operation}. For brevity, the corresponding details of a \textit{scale-down process} is deferred to Appendix~\ref{scale-down-operation}.

\subsection{Initialization}

The initialization phase prepares both the HMM and IMM for fast, low-overhead scaling while ensuring the system can begin serving requests immediately.

Upon startup, the HMM loads the model weights and KV caches for the initial configuration from disk, persists them in device memory, and exposes them through the \texttt{zero\_copy} mechanism. The IMM then instantiates the active inference instance, which attaches to these tensors, and once connected begins receiving queries from the Coordinator and serving them. In addition, the IMM can pre-initialize a set of lightweight standby instances (kept in CPU memory only) for anticipated configurations. These instances are tracked in an LRU cache and remain ready to attach to the HMM’s weights and KV caches on demand, saving significant initialization time during cold starting instances (see Fig.~\ref{fig:init_breakdown} for details).

To this end, each inference instance performs one-time setup tasks such as initializing workers, creating communication groups, and any preparations to bind to the HMM-managed memory state. Only the active instance actually binds to HMM and receives requests from the Coordinator, while suspended instances wait in a ready-to-attach state, minimizing overhead yet allowing fast transition during scale-up or scale-down.

\begin{figure}[htbp]
  \centering
  \includegraphics[width=0.95\linewidth]{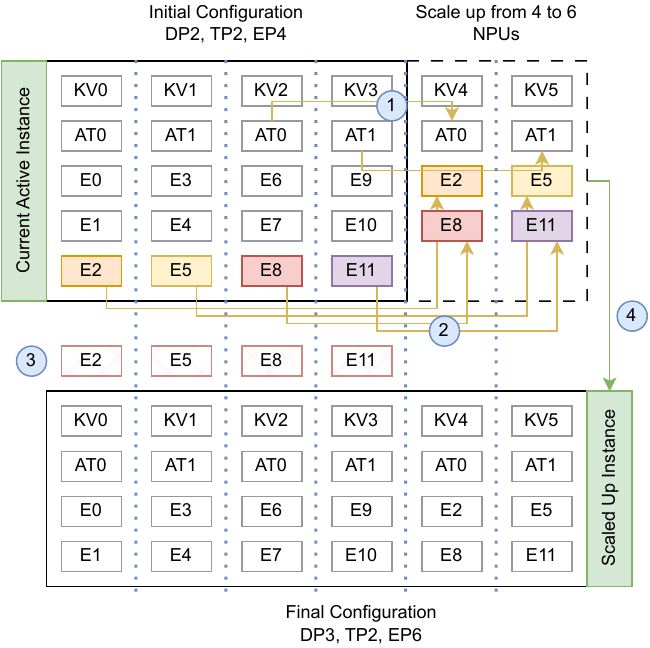}
  \caption{Scale-up lifecycle of HMM during reconfiguration from 4 to 6 NPUs. While the initial instance continues serving, (1) attention weights are copied to the new NPUs via HCCL P2P transfer (yellow arrows), (2) expert weights are migrated to the new NPUs, (3) redundant experts on old NPUs are removed, and (4) weights are zero-copied to activate the scaled-up instance.}
  \label{fig:scale-up-process}
\end{figure}

\subsection{Scale Up Operation}

A scale-up operation in ElasticMoE dynamically expands the set of NPUs allocated to an active inference instance, allowing the system to meet higher request loads without downtime. This operation is jointly orchestrated by the Coordinator, the HMM, and the IMM, each contributing to a staged transition that ensures correctness, performance, and memory efficiency.

Consider a running instance using NPUs \texttt{0--3} under a configuration of DP2-TP2-EP4, and a scale-up request transitions the system to use NPUs \texttt{0--5} with a configuration of DP3-TP2-EP6. This process unfolds in the following steps:

\paragraph{Scale Command Issued}

The process begins when the Coordinator receives a scale-up command, either triggered manually or via autoscaling logic in response to load metrics. It identifies the target configuration and initiates the scaling process by signaling both the HMM and IMM.

\paragraph{HMM Reconfigures Memory Layout}

Upon receiving the scale-up trigger, the HMM first analyzes the current and target configurations to generate a minimal cost plan for weight redistribution. The objective is to maximize zero-copy reuse of existing weights and KV caches, while minimizing the relatively slower P2P transfers. The process is illustrated in Fig.~\ref{fig:scale-up-process}.

\textit{Attention Weights:}  
Since we keep the TP degree fixed during scaling, the attention layers retain their structural layout. This design choice ensures that the attention weights already residing on NPUs \texttt{0--3} can be directly reused by the new inference instance via the \texttt{zero-copy} primitive. For newly added NPUs \texttt{4--5}, attention weights are transferred in a P2P manner from NPUs 0-1 using the \texttt{p2p\_copy} primitive. 

\textit{KV Cache:}  
Similarly, the KV cache structure is also preserved across scale-up, due to the fixed TP configuration. This allows the new inference instance to obtain the existing KV cache directly from NPUs \texttt{0--3}, without reinitialization. Crucially, this prevents memory duplication and avoids allocation spikes during the transition. While the new instance is being prepared, the original instance continues serving in-flight requests using the same existing KV cache. Once the new instance becomes active, it begins serving using this shared cache, allowing seamless handoff and zero service disruption. For NPUs 4,5, KV cache is initialized from scratch.

\textit{Expert Weights:}
When the EP degree changes during scaling, ElasticMoE performs a global remapping of experts to balance placement across NPUs while minimizing data transfer. Expert migration uses \texttt{p2p\_copy} for transferring weights to target devices and \texttt{vpage\_remap} to update virtual–physical mappings in place, eliminating the need to reallocate large contiguous buffers. Old mappings remain active on source devices until the new inference instance takes over, avoiding peak memory bloat and ensuring uninterrupted service.

\paragraph{IMM Prepares New Inference Instance}

While the HMM is finalizing weight configuration, the IMM retrieves the corresponding inference instance from its LRU cache. If the instance does not yet exist (e.g., a \texttt{6-NPU} configuration), it is booted in parallel and stored in the cache for future reuse. 

The IMM prepares this instance up to the point of readiness but defers any weight loading or KV cache allocation. Instead, once the HMM signals readiness, the IMM performs zero-copy attachment of all shared weights and KV caches. At this point, the new inference instance is fully active and ready to serve.

\paragraph{Coordinator Performs Instance Switchover}

Once both the HMM and IMM report that the scaled instance is fully initialized and ready, the Coordinator finalizes the handover. It stops routing new requests to the old instance but allows in-flight requests to finish execution. As soon as the old instance completes all pending queries, it is marked inactive, and all future requests are directed to the newly scaled-up instance.

This switchover is seamless and incurs zero downtime from the perspective of end users. Because both the old and new inference instances share the same memory layout (via zero-copy) and the same KV cache, request latency and batch sizes remain stable throughout the transition.

\section{Implementation}

We implement \textit{ElasticMoE} as three separate modules, Coordinator, IMM, and HMM, that communicate over inter-process communication (IPC) using UNIX domain sockets via the ZMQ framework. The \textit{Coordinator} exposes TCP APIs for receiving user requests and forwards them to the active inference instance in the IMM using standard OpenAI-style inference APIs. The \textit{IMM} manages multiple inference instances, implemented as almost-unmodified vLLM instances, ensuring compatibility with a wide variety of MoE models. 

Within the \textit{HMM}, the control plane is implemented in Python, while the data plane is written in C++ using Huawei CANN APIs for performance-critical operations. These C++ primitives (e.g., zero-copy sharing, peer-to-peer transfers, virtual page remapping) are exposed to Python via PyBind11. To maximize compatibility, the HMM also supports the vLLM model loader backend, allowing all models supported by vLLM to be seamlessly integrated into ElasticMoE.

\section{Evaluation}
We evaluate ElasticMoE against state-of-the-art baselines along five dimensions. First, we measure scaling efficiency (§~\ref{sec:scaling-efficiency}). Second, we analyze SLO recovery after scaling (§~\ref{sec:performance-analysis}). Third, we evaluate SLO attainment under varying request rates (§~\ref{sec:slo-analysis}). Finally, we perform ablations to quantify component contributions (§~\ref{sec:ablations}). Additional experiments, including throughput during scaling (§~\ref{sec:throughput-analysis}) and scale-down latency (§~\ref{sec:scaledown-analysis}), are provided in the appendix.

\subsection{Experimental Setup}
We evaluate the performance of our system under a controlled synthetic workload. The synthetic dataset consists of fixed-length input/output (IO) sequences, enabling deterministic evaluation of scaling behavior. This setup allows for precise measurement of system responsiveness and resource utilization under repeatable conditions. Experiments are conducted in an online (SLO-focused) and offline (throughput-focused) modes. To simulate diverse production-like scenarios, we vary request rates across fixed, variable, and patterned load profiles.

For our experiments, we used the Huawei CloudMatrix384 supernode \cite{zuo2025serving}, which integrates 384 Ascend 910C accelerators and 192 Kunpeng CPUs across 24 nodes. Each node contains 16 Ascend 910C accelerators (64 GB HBM each) and 4 Kunpeng 920 CPUs with 1.5 TB of system RAM. All CPUs and accelerators are interconnected through the Unified Bus (UB), an ultra-high-bandwidth peer-to-peer fabric that offers non-blocking, all-to-all connectivity with near-uniform intra-node and inter-node communication. This tightly coupled design allows CloudMatrix384 to operate as a single large-scale logical node, enabling efficient large-model inference with fine-grained parallelism strategies such as TP and EP.

\subsection{Models and Baselines}

We conduct experiments using three state-of-the-art language models: DeepSeekV2 Lite, a 16B-parameter Mixture-of-Experts (MoE) model with 64 routed experts and 6 activated per token; Qwen3-30B-A3B, a 30.5B-parameter MoE model with 128 experts and 8 activated per token; and DeepSeek V3, a 671B-parameter MoE model with 256 routed experts per layer, of which 8 are activated per token, designed for efficient inference and enhanced reasoning capabilities.

Further, we compare \textit{ElasticMoE} against four baselines, implemented on top of \texttt{vLLM} \cite{kwon2023efficient}. To cover both scaling paradigms, we include one horizontal and four vertical approaches.

\begin{itemize}
\item \textit{Horizontal (Replica):} Launches a new instance as an independent replica on additional accelerators. The old instance continues serving while the new one initializes, ensuring no downtime. However, scaling occurs only in fixed quanta, so even the smallest scaling step effectively doubles the accelerator count.

\item \textit{Vertical (Cold Restart):} Stops the old instance and restarts a new one with the expanded configuration. For example, scaling from 4 to 6 requires exactly 6 accelerators in the final setup, but the system incurs downtime while the old instance is terminated and the new one initializes.

\item \textit{Vertical (Extravagant):} Starts the new instance on fresh accelerators in parallel with the old one. For example, scaling from 4 to 6 requires 10 accelerators in total (4 old + 6 new). This avoids downtime but increases cost by reserving extra accelerators.

\item \textit{Vertical (Colocated):} Starts the new instance on the same accelerators as the old one. For example, when scaling from 4 to 6, the new instance launches on 6 accelerators but reuses the same 4. During scaling, those 4 accelerators must temporarily host two copies of model weights and KV caches, creating high peak memory pressure. To prevent OOM, the KV cache must be reduced in advance, degrading performance.

\end{itemize}

\begin{figure*}[t]
  \centering
  \includegraphics[width=\linewidth]{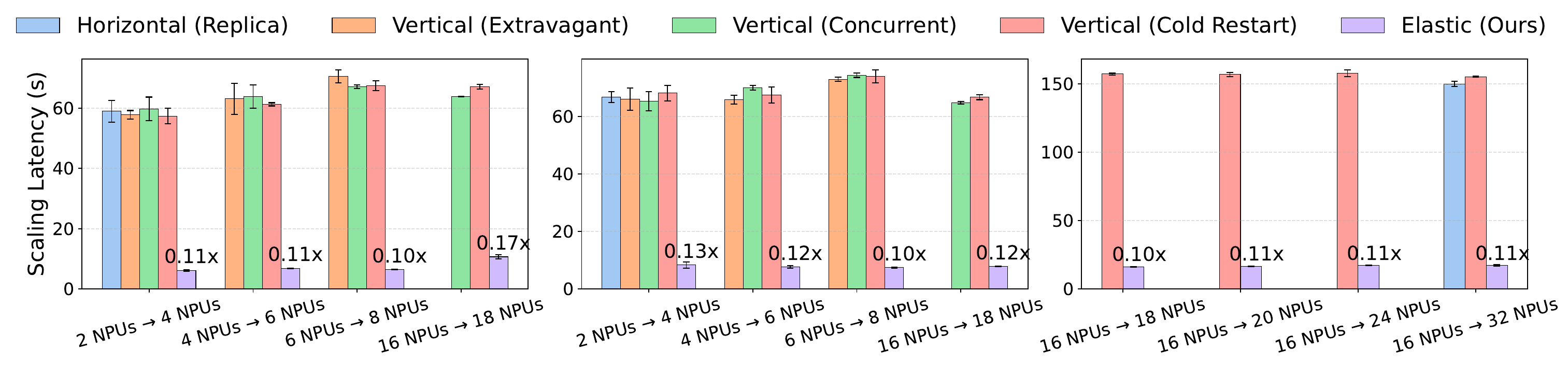}

  \vspace{-0.1em}
  \small (a) DeepSeek V2 Lite \hspace{3.5cm} (b) Qwen 30B-A3B \hspace{3.5cm} (c) DeepSeek V3
  \vspace{0.2em}
  \caption{Scale-up latency comparison across baseline methods for three MoE models. The x-axis indicates scaling configurations, represented as source → destination NPU transitions, corresponding to fixed step size for (a) and (b), and progressively larger steps for (c). In all cases, ElasticMoE consistently achieves substantially lower latency than competing baselines.}
  
  \label{fig:scaling-latency}
\end{figure*}

\subsection{Metrics}
In order to comprehensively compare various scale-up approaches, we test various baselines on two classes of metrics.

\vspace{0.2cm}
\noindent
\textbf{Scaling Metrics:} These metrics capture the responsiveness and overhead of the system as it reacts to changing load.
\begin{itemize}
\item \textit{Scaling Latency:} Time taken from the scale-up command being issued to the new instance being ready to serve.
\item \textit{Downtime:} Interval during scaling when no inference instance (old or new) was available to serve requests.
\item \textit{Peak Memory Usage:} Maximum memory (across all involved NPUs) used during a scaling event.
\end{itemize}

\noindent
\textbf{Performance Metrics:} These metrics evaluate the overall serving quality and efficiency of the system post-scaleup.
\begin{itemize}
\item \textit{TTFT (Time-To-First-Token)}: The elapsed time between a request being submitted to the system and the delivery of the first output token to the user.
\item \textit{TPOT (Time-Per-Output-Token)}: The average time taken to generate each output token, excluding the first token.
\item \textit{SLO Attainment}: The proportion of requests that satisfy predefined SLOs, such as thresholds on TTFT and TPOT, for example TTFT $<$ $\alpha$ and TPOT $<$ $\beta$.
\item \textit{SLO / XPU:} Proportion of requests meeting SLO latency at a fixed RPS, normalized by the number of accelerators.
\end{itemize}

\begin{figure}[h]
\centering
\includegraphics[width=0.95\linewidth]{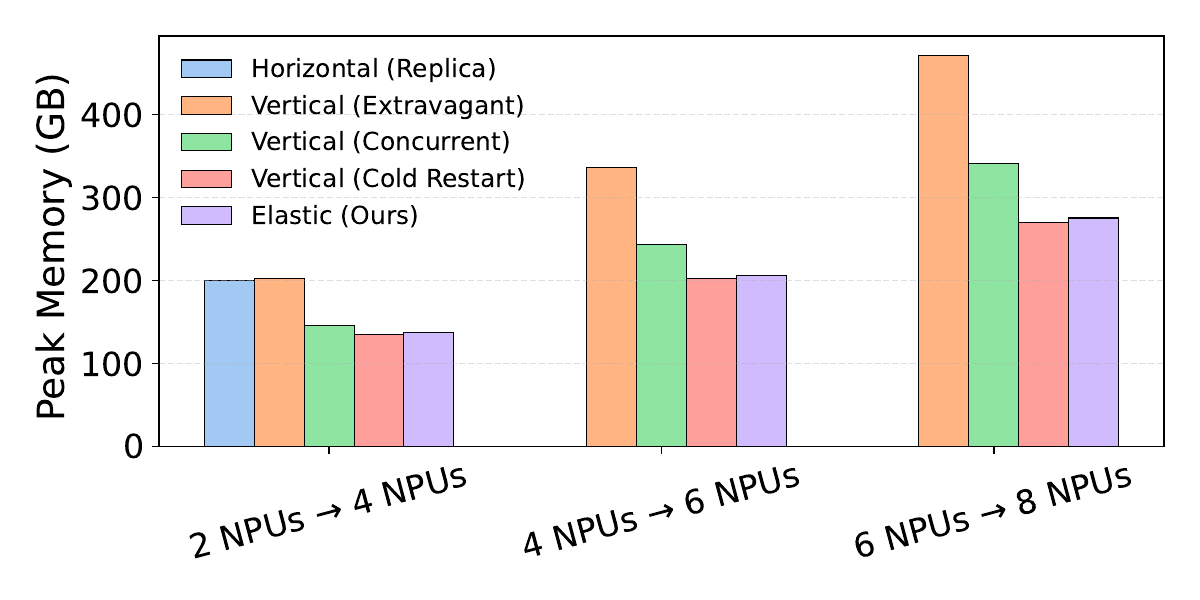}
\caption{Scale-up peak memory across methods for DeepSeek V2 Lite.}
\label{fig:peak-memory}
\vspace{-0.5cm}
\end{figure}

\subsection{Scaling Efficiency} \label{sec:scaling-efficiency}
We evaluate the efficiency of \textit{ElasticMoE} in scale-up operations by measuring scaling latency across different configurations. For DeepSeek V2 Lite and Qwen 30B-A3B, each step corresponds to a 2-NPU increase (e.g., 2$\rightarrow$4, 4$\rightarrow$6). For DeepSeek V3, we also consider larger jumps of 2, 4, 8, and 16 NPUs to study scaling under more aggressive expansions.

Figure~\ref{fig:scaling-latency} summarizes the results. Subfigures (a)--(c) show scale-up latency for DeepSeek V2 Lite, Qwen 30B-A3B, and DeepSeek V3, respectively. The x-axis denotes source $\rightarrow$ destination NPU transitions, while the y-axis reports the time to complete the transition. Notice that, only baselines that are feasible under each configuration are included: the \textit{Extravagant} baseline requires source+target NPUs and is omitted when exceeding available hardware, and the \textit{Horizontal} baseline is feasible only when resources are doubled.

Across all settings, \textit{ElasticMoE} consistently achieves much lower latency than competing methods. Its scale-up latency is only $\approx0.11\times$ that of the best-performing baseline, yielding an improvement of approximately $80.9\%$. These gains arise from ElasticMoE’s design, which combines pre-initialization of inference instances that eliminate cold-start overhead, along with zero-copy sharing of weights and KV caches, and fast peer-to-peer transfers that avoid redundant reloading. This allows scaling to complete rapidly while avoiding the cold restarts or redundant weight loading that dominate baseline costs.

Finally, Fig.~\ref{fig:peak-memory} shows peak memory usage during scale-up on DeepSeek V2 Lite. \textit{Horizontal} and \textit{Vertical (Extravagant)} incur the highest footprints, since they allocate a full new instance in parallel with the old one. \textit{Vertical (Cold Restart)} achieves the lowest as it tears down the old instance before starting the new one. ElasticMoE closely matches Cold Restart, only 2–3\% higher due to live reconfiguration, yet avoids downtime. Compared to Extravagant scaling, it cuts peak memory by 35–40\%.

\subsection{Performance Analysis} \label{sec:performance-analysis}

\begin{figure}[h]
  \centering
  \includegraphics[width=\linewidth]{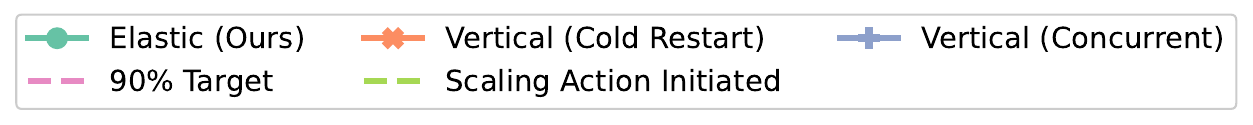}
  \begin{subfigure}[t]{\linewidth}
    \includegraphics[width=1\linewidth]{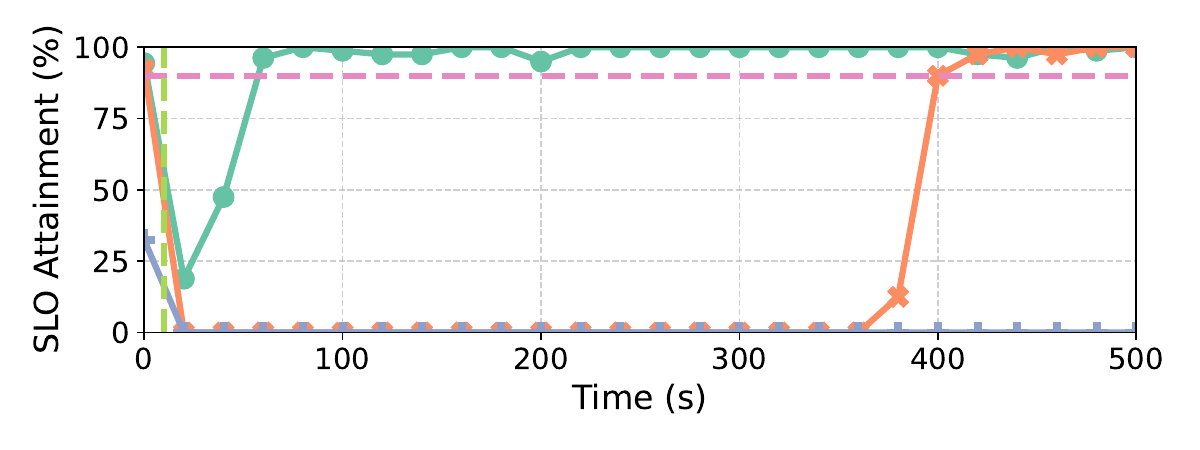}
    \caption{Scale-up from 4$\rightarrow$6 NPUs. Under rising load, all methods initially drop, but ElasticMoE recovers quickly and sustains compliance.}
  \label{fig:scaleup}
  \end{subfigure}
    \begin{subfigure}[t]{\linewidth}
      \includegraphics[width=1\linewidth]{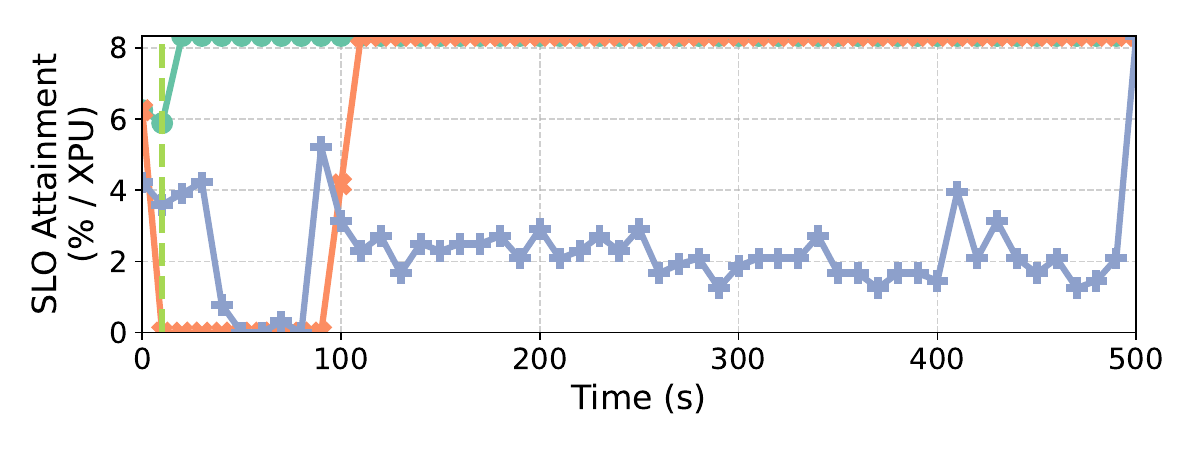}
  \caption{Scale-down from 6$\rightarrow$4 NPUs. With reduced load, ElasticMoE achieves the best normalized SLO attainment by scaling down rapidly and preserving cost efficiency, unlike baselines.}
  \label{fig:scaledown}
    \end{subfigure}
    \caption{SLO dynamics on \textit{DeepSeek V2 Lite}. At $t=0$, workload shifts make the current configuration unsustainable, triggering a scaling action (vertical dotted line).}
  \label{fig:scaling}
\end{figure}

We now evaluate how different methods respond to autoscaling events by tracking SLO dynamics over time. Experiments are conducted on the \textit{DeepSeek V2 Lite} model under synthetic workloads designed to induce scaling actions. At $t=0$, we increase or decrease request load such that the current configuration becomes unsustainable, forcing a scale-up or scale-down decision. Figure~\ref{fig:scaling} reports results, with subfigures (a) and (b) corresponding to scale-up (4$\rightarrow$6 NPUs, target TTFT $\leq$ 5.0s and TPOT $\leq$ 1.5s) and scale-down (6$\rightarrow$4 NPUs, target TTFT $\leq$ 2.0s and TPOT $\leq$ 1.0s), respectively. The vertical dotted line indicates when the scaling command is issued simultaneously across methods. Only relevant baselines are included and infeasible ones (e.g., horizontal) are omitted.

In Fig.~\ref{fig:scaleup}, all methods initially suffer degraded SLO attainment under rising load. ElasticMoE, however, recovers almost immediately after the scaling trigger and sustains compliance above the 90\% target. In Fig.~\ref{fig:scaledown}, the workload decreases, and the system scales from 6$\rightarrow$4 NPUs. Since overall demand is lower, all methods eventually meet SLO requirements. The key difference lies in cost efficiency, measured as normalized SLO attainment per NPU. ElasticMoE maintains high compliance while releasing resources almost immediately after the command, achieving the best SLO-per-NPU.

Overall, ElasticMoE scales up quickly to restore compliance under rising load and scales down smoothly to cut costs under lighter load, all without downtime. In contrast, baselines either suffer from degraded performance or waste resources. ElasticMoE's advantage stems from rapid response due to zero downtime and low scaling latency. By contrast, \textit{Vertical (Cold Restart)} incurs long outages as the old instance is torn down before the new one initializes. \textit{Vertical (Concurrent)} avoids downtime but remains unstable due to overlapping configurations that strain memory and reduce throughput.

\subsection{SLO Compliance Analysis} \label{sec:slo-analysis}
To assess each system’s ability to maintain SLOs under increasing load, we use the \textit{DeepSeek V2 Lite} model with a synthetic workload where RPS grows over time as $rps(t) = f(t)$, simulating realistic traffic patterns. SLO thresholds are fixed (TTFT $\leq$ 1000 ms, TPOT $\leq$ 1000 ms), and all baselines begin with identical resources. A scale-up command is issued at a fixed time to emulate reactive autoscaling. Horizontal scaling is excluded due to infeasibility in this setup. The synthetic workload ensures deterministic behavior, with prompts of ~2000 tokens and decode lengths randomly sampled between 500–750 tokens. This experiment reveals each system’s resource efficiency and scaling responsiveness as load increases.

\begin{figure}[h]
  \centering
  \includegraphics[width=1\linewidth]{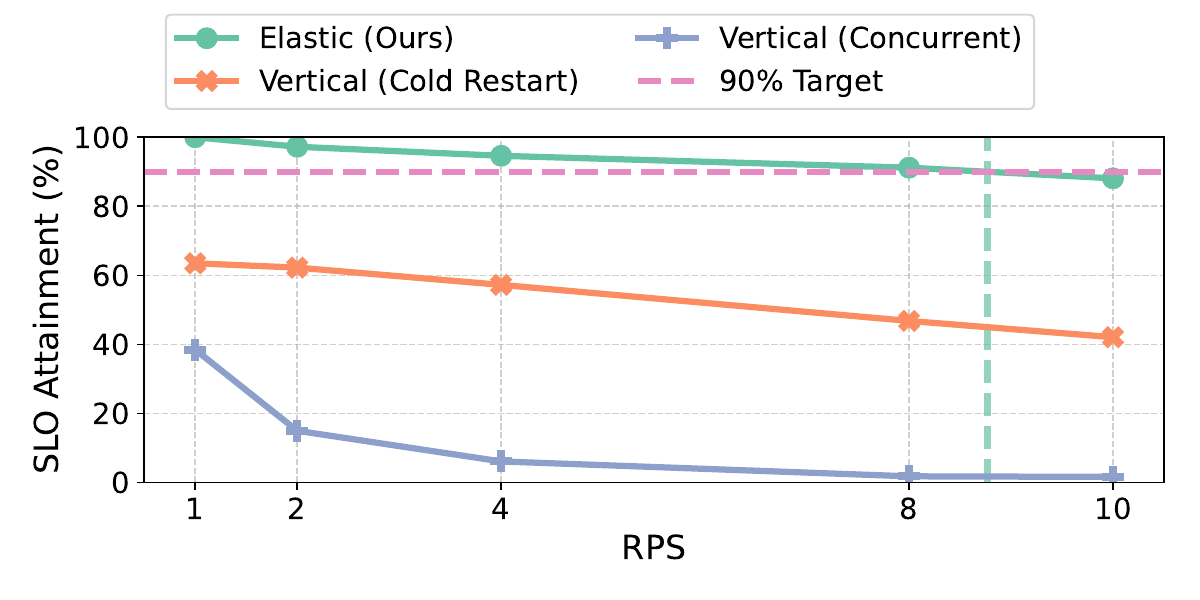}
  \caption{SLO compliance across increasing RPS levels for DeepSeek V2 Lite with a target TTFT $\le$ 1000 ms, TPOT $\le$ 1000 ms. Our method sustains higher SLO\% across load conditions compared to other baselines.}
  \label{fig:slo_rps}
\end{figure}

Figure~\ref{fig:slo_rps} reports the percentage of requests meeting SLOs (y-axis) as the RPS increases (x-axis). Our method consistently maintains compliance above the 90\% threshold up to $\sim$8.7 RPS, demonstrating both high goodput and robustness under rising load. In contrast, \textit{Naive Cold Start} degrades steadily as the load increases, while \textit{Concurrent Vertical} collapses almost immediately, with compliance dropping below 40\% at just 1 RPS and approaching zero as load grows. These results reaffirm earlier conclusions: Naive Cold Start incurs downtime, and Concurrent Vertical sacrifices throughput due to memory constraints. In contrast, our approach eliminates both issues, achieving markedly higher SLOs across all loads.

\begin{table}[t]
\centering
\setlength{\tabcolsep}{5pt} 
\caption{Progressive ablation study of ElasticMoE on scaling from DP3$\rightarrow$DP4. 
Components are disabled cumulatively from top to bottom: first IPC-safe allocator, then HCCL P2P copy, then pre-initialization, and finally zero-copy reuse. 
We report average results over 3 runs on Ascend 910C.}
\label{tab:ablation}
\begin{adjustbox}{max width=.95\linewidth}
\begin{tabular}{lccc}
\toprule
\textbf{Configuration} & 
\makecell{\textbf{Scale}\\\textbf{Time (s)}} & 
\makecell{\textbf{Down}\\\textbf{Time (s)}} & 
\makecell{\textbf{Peak}\\\textbf{Mem. (GB)}} \\
\midrule
ElasticMoE (full) & 2.43 $\pm$ 0.10 & 0 & 275.2 \\
– IPCAlloc        & 3.14 $\pm$ 0.21 & 0 & 290.0 \\
 \hspace{0.1cm} – HCCL            & 10.42 $\pm$ 1.03 & 0 & 290.0 \\
   \hspace{0.2cm} – PreInit         & 62.78 $\pm$ 1.82 & 0 & 290.0 \\
  \hspace{0.3cm}  – ZeroCopy        & 67.40 $\pm$ 1.65 & 67.40 $\pm$ 1.65 & 290.0 \\
\bottomrule
\end{tabular}
\end{adjustbox}
\end{table}

\subsection{Ablation Analysis} \label{sec:ablations}

To quantify the contribution of each design choice, we progressively disable ElasticMoE components (Table~\ref{tab:ablation}). We report scale time, downtime, and peak memory for a scale-up from DP3$\rightarrow$DP4; the corresponding results for scale-down event appear in Appendix~\ref{sec:ablation-scale-down}.

The progression reveals three insights. Removing the IPC-safe allocator slightly increases latency but see a significant raise in peak memory. Disabling HCCL P2P transfers causes an order-of-magnitude slowdown, confirming their importance for fast device provisioning. Eliminating pre-initialization or zero-copy reuse further degrades performance: scale-up latency exceeds 60s, and without zero-copy, downtime is introduced as weights and KV caches must be duplicated complicating reuse. 

Overall, ElasticMoE’s efficiency and zero-downtime scaling rely on the combined effect of memory-efficient allocation, high-bandwidth P2P transfers, pre-initialization, and zero-copy reuse.

\begin{figure}[h]
\centering
\includegraphics[width=0.95\linewidth]{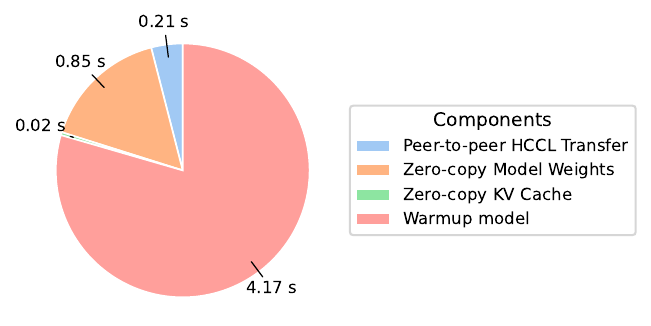}
\caption{Latency breakdown of ElasticMoE scale-up (Qwen 30B-A3B, 12→16 NPUs). Warmup dominates total time, while data movement and zero-copy reuse add negligible overhead.} 
\label{fig:latency-breakdown}
\end{figure}

We now present the latency breakdown of ElasticMoE during scale-up, shown in Fig.~\ref{fig:latency-breakdown}.  The majority of time is spent in model warmup ($\approx$4.2s), whereas P2P transfers, zero-copy weight mapping, and KV-cache reuse only account for a couple of seconds in total. This indicates that the core reconfiguration mechanisms impose minimal overhead. In practice, we assume the target configuration has already been pre-initialized by the IMM, which can anticipate demand and preload nearby configurations. Under this assumption, warmup becomes the dominant cost. If pre-initialization is not available, additional time for full instance preinitialization (on CPU) must be included, which can be significant, as highlighted earlier in Fig.~\ref{fig:init_breakdown}.

\section{Related Work}
We briefly overview the state-of-the-art approaches to automatically scaling LLMs. In addition, we discuss their limitations and compare them with the proposed solution.

\subsection{Instance Replication and Scaling}

Many current works incorporate horizontal scaling~\cite{jin2024p, hu2025deepflow, stojkovic2025dynamollm}, which involves adding or removing entire LLM serving instances to adapt to workload characteristics. This method benefits from simplicity, built-in fault tolerance, and isolation. In contrast, vertical scaling~\cite{stojkovic2025dynamollm, wu2025unlockpotentialfinegrainedllm} adjusts resource allocation to a single instance, which grants fine-grained scaling at the cost of implementation complexity and service downtime. CoCoServe~\cite{wu2025unlockpotentialfinegrainedllm} proposes a vertical scaling approach by replicating a subset of decoder layers onto underutilized hardware. Their work incurs some downtime during scaling and does not inherently support TP or EP, which are required to support larger models that cannot be stored on a single compute device. In contrast, we scale all layers, offer zero downtime during scaling, and support TP and EP.

\subsection{Efficient Scaling}
A line of work~\cite{zhang2025blitzscalefastlivelarge, miao2024spotserve, jeong2023fast, bai2020pipeswitch, hu2025deepflow} has aimed at optimizing resource utilization and minimizing cold start latency. For instance, SpotServe~\cite{miao2024spotserve} and Llumnix~\cite{sun2024llumnix} reduce scaling overhead by lowering the cost of task migration across instances. BlitzScale~\cite{zhang2025blitzscalefastlivelarge} and $\lambda$Scale~\cite{yu2025lambda} leverage high-speed networks between GPUs to multicast model weights when adding new model instances to improve parameter loading times. Tetris~\cite{tetris} identifies tensor redundancies in serverless inferencing and propose allowing function instances to share model weights and tensors, which improves memory efficiency. Despite these, scaling time remains a challenge, and many practical deployments rely on overprovisioning~\cite{zhang2019mark}, which ensures responsiveness but significantly increases operating costs.

\subsection{MoE-Specific Scalability}
A few complementary lines of work specifically address MoE models at scale. The first of these addresses the problem of suboptimal expert placement across devices, which results in poor MoE performance due to imbalanced load over the devices. MoEShard~\cite{Dhasade_2025} is an inference system that achieves optimal load balance by sharding MoE experts. Other works~\cite{wu2024lazarusresilientelastictraining, deepseekai2025deepseekv3technicalreport} look at replicating experts onto hardware based on their usage to better balance workload at the cost of additional memory. 

Another recent topic involves disaggregating the attention modules from the MoE~\cite{zhu2025megascaleinferservingmixtureofexpertsscale,stepfun2025step3largeaffordablemodelsystem, xiao2025xdeepserve}, which allows independent scaling of each. However, to maximize performance and utilization, the modules need to be scaled in specific ratios. In addition, they do not describe any autoscaling functionality.

\section{Conclusion}

This paper presented \textit{ElasticMoE}, a system that enables fine-grained, low-latency, and zero-downtime vertical scaling for large MoE models in production. By decoupling memory from execution, repurposing weights and KV caches via zero-copy and P2P mechanisms, and supporting efficient expert redistribution, ElasticMoE achieves rapid scaling with low peak memory overhead.
Central to this design are the HMM, which manages weights and KV caches independently of inference instances, and the IMM, which anticipates scaling needs through pre-initialization of target configurations. Together, these components enable live reconfiguration without service interruption.
Evaluation shows substantial improvements in scale-up latency, throughput stability, and resource efficiency compared to baselines. Lastly, we discuss the limitations and future work in Appendix.~\ref{sec:limitations-future-work}.

\bibliographystyle{plain}
\bibliography{bibliography}

\begin{thebibliography}{10}

\bibitem{ascend2025ipc}
Ascend cann ipc mechanism api reference.
\newblock \url{https://www.hiascend.com/document/detail/zh/canncommercial/80RC2/apiref/appdevgapi/aclcppdevg_03_0117.html}, 2025.
\newblock Accessed: 2025-09-25.

\bibitem{ascend2025hccl}
Ascend hccl: Huawei collective communication library.
\newblock \url{https://gitee.com/ascend/cann-hccl}, 2025.
\newblock Accessed: 2025-09-25.

\bibitem{bai2020pipeswitch}
Zhihao Bai, Zhen Zhang, Yibo Zhu, and Xin Jin.
\newblock Pipeswitch: Fast pipelined context switching for deep learning applications.
\newblock In {\em 14th USENIX Symposium on Operating Systems Design and Implementation (OSDI 20)}, pages 499--514, 2020.

\bibitem{deepseekai2025deepseekv3technicalreport}
DeepSeek-AI, Aixin Liu, Bei Feng, Bing Xue, Bingxuan Wang, Bochao Wu, Chengda Lu, Chenggang Zhao, Chengqi Deng, Chenyu Zhang, Chong Ruan, Damai Dai, Daya Guo, Dejian Yang, Deli Chen, Dongjie Ji, Erhang Li, Fangyun Lin, Fucong Dai, Fuli Luo, Guangbo Hao, Guanting Chen, Guowei Li, H.~Zhang, Han Bao, Hanwei Xu, Haocheng Wang, Haowei Zhang, Honghui Ding, Huajian Xin, Huazuo Gao, Hui Li, Hui Qu, J.~L. Cai, Jian Liang, Jianzhong Guo, Jiaqi Ni, Jiashi Li, Jiawei Wang, Jin Chen, Jingchang Chen, Jingyang Yuan, Junjie Qiu, Junlong Li, Junxiao Song, Kai Dong, Kai Hu, Kaige Gao, Kang Guan, Kexin Huang, Kuai Yu, Lean Wang, Lecong Zhang, Lei Xu, Leyi Xia, Liang Zhao, Litong Wang, Liyue Zhang, Meng Li, Miaojun Wang, Mingchuan Zhang, Minghua Zhang, Minghui Tang, Mingming Li, Ning Tian, Panpan Huang, Peiyi Wang, Peng Zhang, Qiancheng Wang, Qihao Zhu, Qinyu Chen, Qiushi Du, R.~J. Chen, R.~L. Jin, Ruiqi Ge, Ruisong Zhang, Ruizhe Pan, Runji Wang, Runxin Xu, Ruoyu Zhang, Ruyi Chen, S.~S. Li, Shanghao Lu, Shangyan Zhou, Shanhuang
  Chen, Shaoqing Wu, Shengfeng Ye, Shengfeng Ye, Shirong Ma, Shiyu Wang, Shuang Zhou, Shuiping Yu, Shunfeng Zhou, Shuting Pan, T.~Wang, Tao Yun, Tian Pei, Tianyu Sun, W.~L. Xiao, Wangding Zeng, Wanjia Zhao, Wei An, Wen Liu, Wenfeng Liang, Wenjun Gao, Wenqin Yu, Wentao Zhang, X.~Q. Li, Xiangyue Jin, Xianzu Wang, Xiao Bi, Xiaodong Liu, Xiaohan Wang, Xiaojin Shen, Xiaokang Chen, Xiaokang Zhang, Xiaosha Chen, Xiaotao Nie, Xiaowen Sun, Xiaoxiang Wang, Xin Cheng, Xin Liu, Xin Xie, Xingchao Liu, Xingkai Yu, Xinnan Song, Xinxia Shan, Xinyi Zhou, Xinyu Yang, Xinyuan Li, Xuecheng Su, Xuheng Lin, Y.~K. Li, Y.~Q. Wang, Y.~X. Wei, Y.~X. Zhu, Yang Zhang, Yanhong Xu, Yanhong Xu, Yanping Huang, Yao Li, Yao Zhao, Yaofeng Sun, Yaohui Li, Yaohui Wang, Yi~Yu, Yi~Zheng, Yichao Zhang, Yifan Shi, Yiliang Xiong, Ying He, Ying Tang, Yishi Piao, Yisong Wang, Yixuan Tan, Yiyang Ma, Yiyuan Liu, Yongqiang Guo, Yu~Wu, Yuan Ou, Yuchen Zhu, Yuduan Wang, Yue Gong, Yuheng Zou, Yujia He, Yukun Zha, Yunfan Xiong, Yunxian Ma, Yuting Yan, Yuxiang
  Luo, Yuxiang You, Yuxuan Liu, Yuyang Zhou, Z.~F. Wu, Z.~Z. Ren, Zehui Ren, Zhangli Sha, Zhe Fu, Zhean Xu, Zhen Huang, Zhen Zhang, Zhenda Xie, Zhengyan Zhang, Zhewen Hao, Zhibin Gou, Zhicheng Ma, Zhigang Yan, Zhihong Shao, Zhipeng Xu, Zhiyu Wu, Zhongyu Zhang, Zhuoshu Li, Zihui Gu, Zijia Zhu, Zijun Liu, Zilin Li, Ziwei Xie, Ziyang Song, Ziyi Gao, and Zizheng Pan.
\newblock Deepseek-v3 technical report, 2025.

\bibitem{Dhasade_2025}
Akash Dhasade, Anne-Marie Kermarrec, Erick Lavoie, Johan Pouwelse, Rishi Sharma, and Martijn de~Vos.
\newblock Practical federated learning without a server.
\newblock In {\em Proceedings of the 5th Workshop on Machine Learning and Systems}, EuroMLSys ’25, page 1–11. ACM, March 2025.

\bibitem{hu2025deepflow}
Junhao Hu, Jiang Xu, Zhixia Liu, Yulong He, Yuetao Chen, Hao Xu, Jiang Liu, Baoquan Zhang, Shining Wan, Gengyuan Dan, et~al.
\newblock Deepflow: Serverless large language model serving at scale.
\newblock {\em arXiv preprint arXiv:2501.14417}, 2025.

\bibitem{jeong2023fast}
Jinwoo Jeong, Seungsu Baek, and Jeongseob Ahn.
\newblock Fast and efficient model serving using multi-gpus with direct-host-access.
\newblock In {\em Proceedings of the Eighteenth European Conference on Computer Systems}, pages 249--265, 2023.

\bibitem{jin2024p}
Yibo Jin, Tao Wang, Huimin Lin, Mingyang Song, Peiyang Li, Yipeng Ma, Yicheng Shan, Zhengfan Yuan, Cailong Li, Yajing Sun, et~al.
\newblock P/d-serve: Serving disaggregated large language model at scale.
\newblock {\em arXiv preprint arXiv:2408.08147}, 2024.

\bibitem{khare2025superserve}
Alind Khare, Dhruv Garg, Sukrit Kalra, Snigdha Grandhi, Ion Stoica, and Alexey Tumanov.
\newblock $\{$SuperServe$\}$:$\{$Fine-Grained$\}$ inference serving for unpredictable workloads.
\newblock In {\em 22nd USENIX Symposium on Networked Systems Design and Implementation (NSDI 25)}, pages 739--758, 2025.

\bibitem{kwon2023efficient}
Woosuk Kwon, Zhuohan Li, Siyuan Zhuang, Ying Sheng, Lianmin Zheng, Cody~Hao Yu, Joseph~E. Gonzalez, Hao Zhang, and Ion Stoica.
\newblock Efficient memory management for large language model serving with pagedattention.
\newblock In {\em Proceedings of the ACM SIGOPS 29th Symposium on Operating Systems Principles}, 2023.

\bibitem{lepikhin2021gshard}
Dmitry Lepikhin, HyoukJoong Lee, Yuanzhong Xu, Dehao Chen, Orhan Firat, Yanping Huang, Maxim Krikun, Noam Shazeer, and Zhifeng Chen.
\newblock {\{}GS{\}}hard: Scaling giant models with conditional computation and automatic sharding.
\newblock In {\em International Conference on Learning Representations}, 2021.

\bibitem{tetris}
Jie Li, Laiping Zhao, Yanan Yang, Kunlin Zhan, and Keqiu Li.
\newblock Tetris: Memory-efficient serverless inference through tensor sharing.
\newblock In {\em 2022 USENIX Annual Technical Conference (USENIX ATC 22)}, Carlsbad, CA, July 2022. USENIX Association.

\bibitem{liu2024deepseek}
Aixin Liu, Bei Feng, Bing Xue, Bingxuan Wang, Bochao Wu, Chengda Lu, Chenggang Zhao, Chengqi Deng, Chenyu Zhang, Chong Ruan, et~al.
\newblock Deepseek-v3 technical report.
\newblock {\em arXiv preprint arXiv:2412.19437}, 2024.

\bibitem{miao2024spotserve}
Xupeng Miao, Chunan Shi, Jiangfei Duan, Xiaoli Xi, Dahua Lin, Bin Cui, and Zhihao Jia.
\newblock Spotserve: Serving generative large language models on preemptible instances.
\newblock In {\em Proceedings of the 29th ACM International Conference on Architectural Support for Programming Languages and Operating Systems, Volume 2}, pages 1112--1127, 2024.

\bibitem{moritz2018ray}
Philipp Moritz, Robert Nishihara, Stephanie Wang, Alexey Tumanov, Richard Liaw, Eric Liang, Melih Elibol, Zongheng Yang, William Paul, Michael~I. Jordan, and Ion Stoica.
\newblock Ray: A distributed framework for emerging {AI} applications.
\newblock In {\em 13th USENIX Symposium on Operating Systems Design and Implementation (OSDI 18)}, pages 561--577, Carlsbad, CA, October 2018. USENIX Association.

\bibitem{prabhu2025vattention}
Ramya Prabhu, Ajay Nayak, Jayashree Mohan, Ramachandran Ramjee, and Ashish Panwar.
\newblock vattention: Dynamic memory management for serving llms without pagedattention.
\newblock In {\em Proceedings of the 30th ACM International Conference on Architectural Support for Programming Languages and Operating Systems, Volume 1}, pages 1133--1150, 2025.

\bibitem{stepfun2025step3largeaffordablemodelsystem}
StepFun, Bin Wang, Bojun Wang, Changyi Wan, Guanzhe Huang, Hanpeng Hu, Haonan Jia, Hao Nie, Mingliang Li, Nuo Chen, Siyu Chen, Song Yuan, Wuxun Xie, Xiaoniu Song, Xing Chen, Xingping Yang, Xuelin Zhang, Yanbo Yu, Yaoyu Wang, Yibo Zhu, Yimin Jiang, Yu~Zhou, Yuanwei Lu, Houyi Li, Jingcheng Hu, Ka~Man Lo, Ailin Huang, Binxing Jiao, Bo~Li, Boyu Chen, Changxin Miao, Chang Lou, Chen Hu, Chen Xu, Chenfeng Yu, Chengyuan Yao, Daokuan Lv, Dapeng Shi, Deshan Sun, Ding Huang, Dingyuan Hu, Dongqing Pang, Enle Liu, Fajie Zhang, Fanqi Wan, Gulin Yan, Han Zhang, Han Zhou, Hanghao Wu, Hangyu Guo, Hanqi Chen, Hanshan Zhang, Hao Wu, Haocheng Zhang, Haolong Yan, Haoran Lv, Haoran Wei, Hebin Zhou, Heng Wang, Heng Wang, Hongxin Li, Hongyu Zhou, Hongyuan Wang, Huiyong Guo, Jia Wang, Jiahao Gong, Jialing Xie, Jian Zhou, Jianjian Sun, Jiaoren Wu, Jiaran Zhang, Jiayu Liu, Jie Cheng, Jie Luo, Jie Yan, Jie Yang, Jieyi Hou, Jinguang Zhang, Jinlan Cao, Jisheng Yin, Junfeng Liu, Junhao Huang, Junzhe Lin, Kaijun Tan, Kaixiang Li, Kang An,
  Kangheng Lin, Kenkun Liu, Lei Yang, Liang Zhao, Liangyu Chen, Lieyu Shi, Liguo Tan, Lin Lin, Lin Zhang, Lina Chen, Liwen Huang, Liying Shi, Longlong Gu, Mei Chen, Mengqiang Ren, Ming Li, Mingzhe Chen, Na~Wang, Nan Wu, Qi~Han, Qian Zhao, Qiang Zhang, Qianni Liu, Qiaohui Chen, Qiling Wu, Qinglin He, Qinyuan Tan, Qiufeng Wang, Qiuping Wu, Qiuyan Liang, Quan Sun, Rui Li, Ruihang Miao, Ruosi Wan, Ruyan Guo, Shangwu Zhong, Shaoliang Pang, Shengjie Fan, Shijie Shang, Shilei Jiang, Shiliang Yang, Shiming Hao, Shuli Gao, Siming Huang, Siqi Liu, Tiancheng Cao, Tianhao Cheng, Tianhao Peng, Wang You, Wei Ji, Wen Sun, Wenjin Deng, Wenqing He, Wenzhen Zheng, Xi~Chen, Xiangwen Kong, Xianzhen Luo, Xiaobo Yang, Xiaojia Liu, Xiaoxiao Ren, Xin Han, Xin Li, Xin Wu, Xu~Zhao, Yanan Wei, Yang Li, Yangguang Li, Yangshijie Xu, Yanming Xu, Yaqiang Shi, Yeqing Shen, Yi~Yang, Yifei Yang, Yifeng Gong, Yihan Chen, Yijing Yang, Yinmin Zhang, Yizhuang Zhou, Yuanhao Ding, Yuantao Fan, Yuanzhen Yang, Yuchu Luo, Yue Peng, Yufan Lu, Yuhang
  Deng, Yuhe Yin, Yujie Liu, Yukun Chen, Yuling Zhao, Yun Mou, Yunlong Li, Yunzhou Ju, Yusheng Li, Yuxiang Yang, Yuxiang Zhang, Yuyang Chen, Zejia Weng, Zhe Xie, Zheng Ge, Zheng Gong, Zhenyi Lu, Zhewei Huang, Zhichao Chang, Zhiguo Huang, Zhirui Wang, Zidong Yang, Zili Wang, Ziqi Wang, Zixin Zhang, Binxing Jiao, Daxin Jiang, Heung-Yeung Shum, and Xiangyu Zhang.
\newblock Step-3 is large yet affordable: Model-system co-design for cost-effective decoding, 2025.

\bibitem{stojkovic2025dynamollm}
Jovan Stojkovic, Chaojie Zhang, {\'I}{\~n}igo Goiri, Josep Torrellas, and Esha Choukse.
\newblock Dynamollm: Designing llm inference clusters for performance and energy efficiency.
\newblock In {\em 2025 IEEE International Symposium on High Performance Computer Architecture (HPCA)}, pages 1348--1362. IEEE, 2025.

\bibitem{sun2024llumnix}
Biao Sun, Ziming Huang, Hanyu Zhao, Wencong Xiao, Xinyi Zhang, Yong Li, and Wei Lin.
\newblock Llumnix: Dynamic scheduling for large language model serving.
\newblock In {\em 18th USENIX Symposium on Operating Systems Design and Implementation (OSDI 24)}, pages 173--191, 2024.

\bibitem{kimiteam2025kimik2openagentic}
Kimi Team, Yifan Bai, Yiping Bao, Guanduo Chen, Jiahao Chen, Ningxin Chen, Ruijue Chen, Yanru Chen, Yuankun Chen, Yutian Chen, Zhuofu Chen, Jialei Cui, Hao Ding, Mengnan Dong, Angang Du, Chenzhuang Du, Dikang Du, Yulun Du, Yu~Fan, Yichen Feng, Kelin Fu, Bofei Gao, Hongcheng Gao, Peizhong Gao, Tong Gao, Xinran Gu, Longyu Guan, Haiqing Guo, Jianhang Guo, Hao Hu, Xiaoru Hao, Tianhong He, Weiran He, Wenyang He, Chao Hong, Yangyang Hu, Zhenxing Hu, Weixiao Huang, Zhiqi Huang, Zihao Huang, Tao Jiang, Zhejun Jiang, Xinyi Jin, Yongsheng Kang, Guokun Lai, Cheng Li, Fang Li, Haoyang Li, Ming Li, Wentao Li, Yanhao Li, Yiwei Li, Zhaowei Li, Zheming Li, Hongzhan Lin, Xiaohan Lin, Zongyu Lin, Chengyin Liu, Chenyu Liu, Hongzhang Liu, Jingyuan Liu, Junqi Liu, Liang Liu, Shaowei Liu, T.~Y. Liu, Tianwei Liu, Weizhou Liu, Yangyang Liu, Yibo Liu, Yiping Liu, Yue Liu, Zhengying Liu, Enzhe Lu, Lijun Lu, Shengling Ma, Xinyu Ma, Yingwei Ma, Shaoguang Mao, Jie Mei, Xin Men, Yibo Miao, Siyuan Pan, Yebo Peng, Ruoyu Qin, Bowen Qu, Zeyu
  Shang, Lidong Shi, Shengyuan Shi, Feifan Song, Jianlin Su, Zhengyuan Su, Xinjie Sun, Flood Sung, Heyi Tang, Jiawen Tao, Qifeng Teng, Chensi Wang, Dinglu Wang, Feng Wang, Haiming Wang, Jianzhou Wang, Jiaxing Wang, Jinhong Wang, Shengjie Wang, Shuyi Wang, Yao Wang, Yejie Wang, Yiqin Wang, Yuxin Wang, Yuzhi Wang, Zhaoji Wang, Zhengtao Wang, Zhexu Wang, Chu Wei, Qianqian Wei, Wenhao Wu, Xingzhe Wu, Yuxin Wu, Chenjun Xiao, Xiaotong Xie, Weimin Xiong, Boyu Xu, Jing Xu, Jinjing Xu, L.~H. Xu, Lin Xu, Suting Xu, Weixin Xu, Xinran Xu, Yangchuan Xu, Ziyao Xu, Junjie Yan, Yuzi Yan, Xiaofei Yang, Ying Yang, Zhen Yang, Zhilin Yang, Zonghan Yang, Haotian Yao, Xingcheng Yao, Wenjie Ye, Zhuorui Ye, Bohong Yin, Longhui Yu, Enming Yuan, Hongbang Yuan, Mengjie Yuan, Haobing Zhan, Dehao Zhang, Hao Zhang, Wanlu Zhang, Xiaobin Zhang, Yangkun Zhang, Yizhi Zhang, Yongting Zhang, Yu~Zhang, Yutao Zhang, Yutong Zhang, Zheng Zhang, Haotian Zhao, Yikai Zhao, Huabin Zheng, Shaojie Zheng, Jianren Zhou, Xinyu Zhou, Zaida Zhou, Zhen Zhu,
  Weiyu Zhuang, and Xinxing Zu.
\newblock Kimi k2: Open agentic intelligence, 2025.

\bibitem{wu2025unlockpotentialfinegrainedllm}
Jingfeng Wu, Yiyuan He, Minxian Xu, Xitong Gao, Kejiang Ye, and Chengzhong Xu.
\newblock Unlock the potential of fine-grained llm serving via dynamic module scaling, 2025.

\bibitem{wu2024lazarusresilientelastictraining}
Yongji Wu, Wenjie Qu, Tianyang Tao, Zhuang Wang, Wei Bai, Zhuohao Li, Yuan Tian, Jiaheng Zhang, Matthew Lentz, and Danyang Zhuo.
\newblock Lazarus: Resilient and elastic training of mixture-of-experts models with adaptive expert placement, 2024.

\bibitem{xiao2025xdeepserve}
Ao~Xiao, Bangzheng He, Baoquan Zhang, Baoxing Huai, Bingji Wang, Bo~Wang, Bo~Xu, Boyi Hou, Chan Yang, Changhong Liu, Cheng Cui, Chenyu Zhu, Cong Feng, Daohui Wang, Dayun Lin, Duo Zhao, Fengshao Zou, Fu~Wang, Gangqiang Zhang, Gengyuan Dan, Guanjie Chen, Guodong Guan, Guodong Yang, Haifeng Li, Haipei Zhu, Haley Li, Hao Feng, Hao Huang, Hao Xu, Hengrui Ma, Hengtao Fan, Hui Liu, Jia Li, Jiang Liu, Jiang Xu, Jie Meng, Jinhan Xin, Junhao Hu, Juwei Chen, Lan Yu, Lanxin Miao, Liang Liu, Linan Jing, Lu~Zhou, Meina Han, Mingkun Deng, Mingyu Deng, Naitian Deng, Nizhong Lin, Peihan Zhao, Peng Pan, Pengfei Shen, Ping Li, Qi~Zhang, Qian Wang, Qin ZhC~Qingrong Xia, Qingyi Zhang, Qunchao Fu, Ren Guo, Ruimin Gao, Shaochun Li, Sheng Long, Shentian Li, Shining Wan, Shuai Shen, Shuangfu Zeng, Shuming Jing, Siqi Yang, Song Zhang, Tao Xu, Tianlin Du, Ting Chen, Wanxu Wu, Wei Jiang, Weinan Tong, Weiwei Chen, Wen Peng, Wenli Zhou, Wenquan Yang, Wenxin Liang, Xiang Liu, Xiaoli Zhou, Xin Jin, Xinyu Duan, Xu~Li, Xu~Zhang, Xusheng Chen,
  Yalong Shan, Yang Gan, Yao Lu, Yi~Deng, Yi~Zheng, Ying Xiong, Yingfei Zheng, Yiyun Zheng, Yizhou Shan, Yong Gao, Yong Zhang, Yongqiang Yang, Yuanjin Gong, Yue Yu, Yuetao Chen, Yukun Zhu, Yulong He, Yusu Zhao, Yuyan Wu, Zenan Zhang, Zhaojin Zhuo, Zhaoyang Ji, Zhefeng Wang, Zheng Wang, Zhenan Fan, Zhenhua Yang, Zhenli Sheng, Zhibin Yu, Zhigang Ji, Zhihao Ren, Zhipeng Bian, Zhixia Liu, Zhiyu Dong, Zhonghua Li, Zhou Yu, Zhuoming Shen, Zhuwei Peng, Zi~Ye, Zihao Xiang, Zimin Fu, and Zixuan Zhang.
\newblock xdeepserve: Model-as-a-service on huawei cloudmatrix384, 2025.

\bibitem{qwen3}
An~Yang, Anfeng Li, Baosong Yang, Beichen Zhang, Binyuan Hui, Bo~Zheng, Bowen Yu, Chang Gao, Chengen Huang, Chenxu Lv, Chujie Zheng, Dayiheng Liu, Fan Zhou, Fei Huang, Feng Hu, Hao Ge, Haoran Wei, Huan Lin, Jialong Tang, Jian Yang, Jianhong Tu, Jianwei Zhang, Jianxin Yang, Jiaxi Yang, Jing Zhou, Jingren Zhou, Junyang Lin, Kai Dang, Keqin Bao, Kexin Yang, Le~Yu, Lianghao Deng, Mei Li, Mingfeng Xue, Mingze Li, Pei Zhang, Peng Wang, Qin Zhu, Rui Men, Ruize Gao, Shixuan Liu, Shuang Luo, Tianhao Li, Tianyi Tang, Wenbiao Yin, Xingzhang Ren, Xinyu Wang, Xinyu Zhang, Xuancheng Ren, Yang Fan, Yang Su, Yichang Zhang, Yinger Zhang, Yu~Wan, Yuqiong Liu, Zekun Wang, Zeyu Cui, Zhenru Zhang, Zhipeng Zhou, and Zihan Qiu.
\newblock Qwen3 technical report.
\newblock {\em arXiv preprint arXiv:2505.09388}, 2025.

\bibitem{qwen2.5}
An~Yang, Baosong Yang, Beichen Zhang, Binyuan Hui, Bo~Zheng, Bowen Yu, Chengyuan Li, Dayiheng Liu, Fei Huang, Haoran Wei, Huan Lin, Jian Yang, Jianhong Tu, Jianwei Zhang, Jianxin Yang, Jiaxi Yang, Jingren Zhou, Junyang Lin, Kai Dang, Keming Lu, Keqin Bao, Kexin Yang, Le~Yu, Mei Li, Mingfeng Xue, Pei Zhang, Qin Zhu, Rui Men, Runji Lin, Tianhao Li, Tingyu Xia, Xingzhang Ren, Xuancheng Ren, Yang Fan, Yang Su, Yichang Zhang, Yu~Wan, Yuqiong Liu, Zeyu Cui, Zhenru Zhang, and Zihan Qiu.
\newblock Qwen2.5 technical report.
\newblock {\em arXiv preprint arXiv:2412.15115}, 2024.

\bibitem{yu2025lambda}
Minchen Yu, Rui Yang, Chaobo Jia, Zhaoyuan Su, Sheng Yao, Tingfeng Lan, Yuchen Yang, Yue Cheng, Wei Wang, Ao~Wang, et~al.
\newblock $\lambda$scale: Enabling fast scaling for serverless large language model inference.
\newblock {\em arXiv preprint arXiv:2502.09922}, 2025.

\bibitem{zhang2019mark}
Chengliang Zhang, Minchen Yu, Wei Wang, and Feng Yan.
\newblock Mark: Exploiting cloud services for cost-effective, slo-aware machine learning inference serving.
\newblock In {\em 2019 USENIX Annual Technical Conference (USENIX ATC 19)}, pages 1049--1062, 2019.

\bibitem{zhang2025blitzscalefastlivelarge}
Dingyan Zhang, Haotian Wang, Yang Liu, Xingda Wei, Yizhou Shan, Rong Chen, and Haibo Chen.
\newblock Blitzscale: Fast and live large model autoscaling with o(1) host caching, 2025.

\bibitem{zhu2025megascaleinferservingmixtureofexpertsscale}
Ruidong Zhu, Ziheng Jiang, Chao Jin, Peng Wu, Cesar~A. Stuardo, Dongyang Wang, Xinlei Zhang, Huaping Zhou, Haoran Wei, Yang Cheng, Jianzhe Xiao, Xinyi Zhang, Lingjun Liu, Haibin Lin, Li-Wen Chang, Jianxi Ye, Xiao Yu, Xuanzhe Liu, Xin Jin, and Xin Liu.
\newblock Megascale-infer: Serving mixture-of-experts at scale with disaggregated expert parallelism, 2025.

\bibitem{zuo2025serving}
Pengfei Zuo, Huimin Lin, Junbo Deng, Nan Zou, Xingkun Yang, Yingyu Diao, Weifeng Gao, Ke~Xu, Zhangyu Chen, Shirui Lu, et~al.
\newblock Serving large language models on huawei cloudmatrix384.
\newblock {\em arXiv preprint arXiv:2506.12708}, 2025.

\end{thebibliography}

\balance 
\clearpage
\appendix
\section{Additional Experiments}

\subsection{Throughput analysis} \label{sec:throughput-analysis}
We evaluate inference throughput during scaling transitions using the \textit{DeepSeek V2 Lite} model in an offline batch processing setting. An offline batch of 10000 requests is drawn from a synthetic workload with 500 prefill tokens and a random range of 250-500 decode tokens. The system is initially provisioned with 6 NPUs and scales up to 8 NPUs at a fixed time for all baselines. Baselines that are not applicable for this transition are excluded. To capture scaling behavior consistently, we divide execution into three windows: \textbf{before} scaling, \textbf{during} scaling, and \textbf{after} scaling. The ``during'' window is measured as $\pm$5 seconds around the longest scaling transition among all baselines (in this case, Vertical Cold Restart). This setup allows us to examine how different methods perform in steady-state, during the critical transition, and after additional capacity becomes available.

\begin{table}[H]
  \centering
  \caption{Throughput comparison (requests/sec) for scale-up from DP3TP2 to DP4TP2. We report throughput for three windows: \textit{before}, {during}, and {after} scaling.}
  \label{tab:throughput_scaleup}
  \begin{tabular}{lccc}
    \toprule
    \textbf{Method} & \textbf{Before} & \textbf{During} & \textbf{After} \\
    \midrule
    Vertical (Concurrent) & 1.338 & 0.467 & 2.268  \\
    Vertical (Cold Restart) & 6.002  & 2.064 & 7.818 \\
    Elastic (Ours)   & \textit{6.002} & \textit{3.943} & \textit{7.818} \\
    \bottomrule
  \end{tabular}
\end{table} 

Table~\ref{tab:throughput_scaleup} reports throughput across the three windows. Before scaling, both ElasticMoE and Cold Restart achieve similar throughput, while Concurrent performs poorly because it reserves memory for potential scaling and thus operates at reduced capacity at all times. During scaling, ElasticMoE sustains the highest throughput—nearly double that of Cold Restart—due to lower scaling latency and zero downtime. Although throughput temporarily dips compared to steady state (since the active instance pauses new request intake, lowering effective batch size), service remains uninterrupted, avoiding downtime entirely. After scaling, all methods benefit from the added NPUs and achieve higher throughput than before.

These results demonstrate that ElasticMoE not only scales faster than baselines but also maintains substantially higher throughput during the critical transition period, combining zero downtime with efficient resource utilization.

\begin{table}[t]
\centering
\setlength{\tabcolsep}{5pt} 
\caption{Progressive ablation study of ElasticMoE on scaling down from DP4$\rightarrow$DP3. 
Components are disabled cumulatively from top to bottom: first IPC-safe allocator, then HCCL P2P copy, then pre-initialization, and finally zero-copy reuse. 
We report average results over 3 runs on Ascend 910C.}
\label{tab:ablation-scaledown}
\begin{adjustbox}{max width=.95\linewidth}
\begin{tabular}{lccc}
\toprule
\textbf{Configuration} & 
\makecell{\textbf{Scale}\\\textbf{Time (s)}} & 
\makecell{\textbf{Down}\\\textbf{Time (s)}} & 
\makecell{\textbf{Peak}\\\textbf{Mem. (GB)}} \\
\midrule
ElasticMoE (full) & 1.38 $\pm$ 0.10 & 0 & 274.4 \\
– {IPCAlloc}        & 1.36 $\pm$ 0.17 & 0 & 290.0 \\
 \hspace{0.1cm} – {HCCL}      & 7.74 $\pm$ 0.62 & 0 & 290.0 \\
 \hspace{0.2cm} – {PreInit}   & 50.21 $\pm$ 0.96 & 0 & 290.0 \\
 \hspace{0.3cm} – {ZeroCopy} & 64.57 $\pm$ 4.28 & 64.57 $\pm$ 4.28 & 290.0 \\
\bottomrule
\end{tabular}
\end{adjustbox}
\end{table}

\subsection{Scale-down Latency Analysis} \label{sec:scaledown-analysis}

\begin{figure*}[t]
  \centering
  \includegraphics[width=\linewidth]{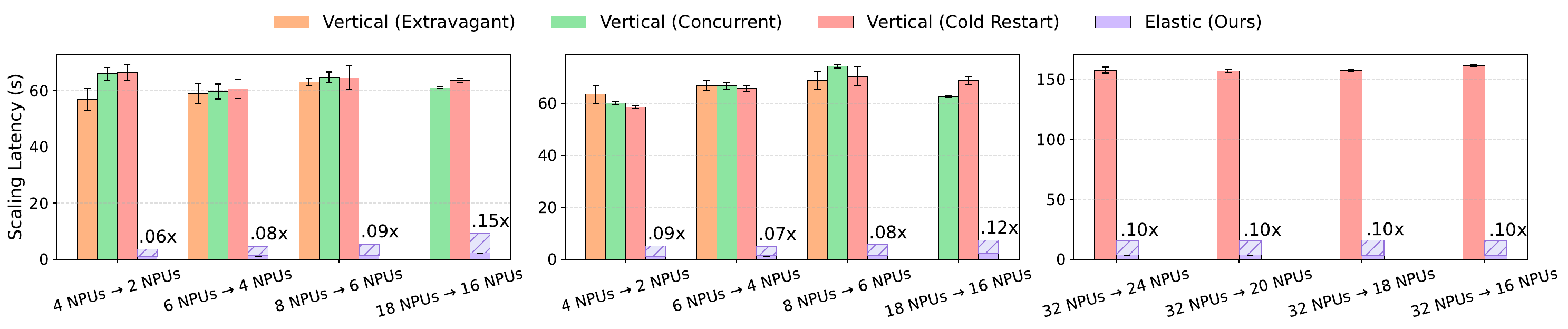}
  \vspace{-0.1em}
  \small (a) DeepSeek V2 Lite \hspace{3.5cm} (b) Qwen 30B-A3B \hspace{3.5cm} (c) DeepSeek V3
  \vspace{0.2em}
\caption{Scale-down latency comparison across baseline methods for three MoE models. The x-axis indicates scaling configurations, represented as source → destination NPU transitions, corresponding to fixed step size for (a) and (b), and progressively larger steps for (c). Similar to the scale-up scenario, in all cases, ElasticMoE consistently achieves substantially lower latency than competing baselines. In our approach, the shaded purple region denotes the warm-up time.}
  
  \label{fig:scaledown-latency}
\end{figure*}

We complement the scale-up evaluation by analyzing scale-down behavior across the same three models. 
Here, scaling transitions reduce the number of NPUs, with step sizes of 2 for DeepSeek V2 Lite and Qwen 30B-A3B, and progressively larger steps for DeepSeek V3. 
Figure~\ref{fig:scaledown-latency} reports the results.

As in the scale-up scenario, \textit{ElasticMoE} achieves substantially lower latency than competing baselines. Across all tested configurations, our method consistently completes scale-down in less than $0.15\times$ the time of the fastest baseline. This translates into latency reductions of 80--90\% relative to conventional vertical scaling methods. \textit{Vertical (Cold Restart)} suffers long downtime since the old instance must terminate before the new instance initializes, while \textit{Vertical (Extravagant)} and \textit{Vertical (Concurrent)} temporarily maintain overlapping configurations, inflating latency and memory usage. In contrast, ElasticMoE reclaims resources immediately via live reconfiguration, avoiding redundant weight reloading and memory spikes. 
The benefits are particularly pronounced in DeepSeek V3, where even aggressive 16$\rightarrow$2 NPU reductions complete with $\approx0.10\times$ baseline latency. 

Overall, these results confirm that ElasticMoE not only scales up quickly to meet rising demand but also scales down efficiently to release capacity, ensuring cost-effective elasticity with minimal disruption.

\subsection{Ablation Analysis for Scale-Down} \label{sec:ablation-scale-down}

Table~\ref{tab:ablation-scaledown} reports the progressive ablation study for a scale-down event (DP4$\rightarrow$DP3). The trends mirror those in scale-up. Disabling the IPC-safe allocator has negligible effect on latency but increases peak memory. Removing HCCL transfers significantly slows the transition. Eliminating pre-initialization and zero-copy reuse further worsens performance: scale-down latency rises above 60s, and without zero-copy, downtime is introduced because weights and KV caches must be duplicated. These results reinforce that all four mechanisms—efficient allocation, P2P transfers, pre-initialization, and zero-copy reuse—are jointly essential for low-latency, zero-downtime scaling in both directions.

\section{Extending to the CUDA Ecosystem}

Although ElasticMoE is implemented on Ascend NPUs, the framework can be readily extended to the CUDA ecosystem. We have developed a barebones proof-of-concept implementation on NVIDIA GPUs that confirms the feasibility of this port.

On the \textit{HMM side}, the control-plane logic remains unchanged, since resource tracking, scaling orchestration, and zero-copy coordination are device-agnostic. The data plane, however, must replace CANN-specific primitives with CUDA equivalents. For example, CUDA provides 
\texttt{cuda\allowbreak Ipc\allowbreak GetMem\allowbreak Handle}, 
\texttt{cudaIpc\allowbreak OpenMem\allowbreak Handle}, and \texttt{cuda\allowbreak Malloc} for inter-process memory sharing, which substitute Ascend’s IPC APIs. Similarly, virtual expert management can be supported through CUDA’s virtual memory primitives \cite{prabhu2025vattention}, such as \texttt{cuMemAddressReserve} and \texttt{cuMemMap}, which enable page-based allocation and remapping.

On the \textit{IMM side}, the design also remains largely unchanged. The same zero-copy loader and instance manager can be reused, with the only difference being the backend: instead of \texttt{ascend-vLLM}, the implementation would rely on the standard \texttt{vLLM} for CUDA-enabled GPUs.

Overall, extending ElasticMoE to CUDA primarily involves swapping low-level device APIs, while the higher-level control, coordination, and inference logic remains intact. This demonstrates that ElasticMoE’s design is portable across accelerator ecosystems with minimal changes.

\section{Limitations and Future Work} \label{sec:limitations-future-work}

While ElasticMoE demonstrates the feasibility and benefits of fine-grained, zero-downtime vertical scaling, several limitations remain.

First, although the system supports fine-grained scaling through adjustments in DP and EP degrees, the TP degree needs to be held fixed. This restriction simplifies migration by keeping shared model weights and KV cache layouts unchanged, but it also constrains the granularity of scaling. For some configurations high TP (although rare), the minimum scaling unit remains tied to TP size, limiting elasticity and granularity.

Second, ElasticMoE eliminates downtime by keeping the active instance serving requests while the new configuration is prepared in parallel. However, during the transition, the active instance pauses intake of new requests, which reduces effective batch size and temporarily lowers throughput (Table~\ref{tab:throughput_scaleup}). While preferable to downtime, this reduced capacity highlights a trade-off between availability and performance during scaling.

\medskip
\noindent\textbf{Future work.}  
Building on the above-mentioned limitations, one promising direction is to relax the fixed-TP constraint. Supporting flexible TP degrees would allow even finer scaling but introduces new challenges, such as complex sharding, migrating weights and reshaping KV caches without incurring high latency or service interruption. Developing techniques to manage this additional complexity while keeping scale-up latency small remains an open systems problem.

Another avenue is to improve transition capacity. Currently, the active instance operates at reduced throughput during scale-up. A more ambitious design would keep the old instance at full capacity until the new one is ready, or even allow both instances to serve requests concurrently sharing compute resources. Realizing this vision would require mechanisms to manage two independent activation spaces, coordinate KV cache block allocation across instances, and safely migrate requests in-flight—posing interesting challenges for distributed memory management and scheduling.

\section{Implementation Details for Low-level Primitives} \label{low-level-premitives-implementation-details}
ElasticMoE implements a set of low-level memory and communication primitives within the HMM and the IMM to enable fine-grained, low-latency scaling with minimal peak memory overhead. These primitives provide the foundational mechanisms for weight allocation, transfer, and sharing across NPUs, allowing the system to efficiently reuse model state and avoid redundant data movement during scale operations.  Specifically, these allow to allocate, distribute, and share model weights and KV caches across NPUs efficiently. These are explained as follows:

\subsection{IPC-Compatible Tensor Allocation (\texttt{IpcSafeAllocator})}

To enable zero-copy memory sharing across processes, ElasticMoE overrides PyTorch’s default memory allocator, \texttt{Torch\allowbreak Caching\allowbreak Allocator}, with a custom allocator designed for IPC-safe memory allocation. While the default allocator uses a device memory pool, the resulting allocations are typically managed as a single block, making them incompatible with inter-process communication (IPC) sharing.

In contrast, our \texttt{Ipc\allowbreak Safe\allowbreak Allocator} directly allocates physical memory regions using hardware-specific APIs that mark the memory as IPC-compatible. We override core PyTorch allocation functions such as \texttt{torch.\allowbreak ones()}, \texttt{torch.\allowbreak empty()} and \texttt{torch.\allowbreak full()} to ensure that all model weights intended for sharing are allocated via this mechanism. This makes tensors accessible across processes and avoids the need for redundant copies during scaling transitions.

\subsection{Direct Disk-to-NPU Weight Loading (\texttt{disk-copy})}
Naively using the accelerator device map functionality to load model weights may lead to same tensors (across different NPUs) being read and loaded from disk. This is suboptimal because the disk-to-NPU transfer is the slowest link in the data path, which typically stages tensors in host memory before moving them to device memory.  
Hence, we implement the \texttt{disk-copy} primitive that can read and load only a subset of tensors—selected by name, partition index (e.g., TP rank), or layer type—from disk to the target NPU. This ensures no tensor is loaded from disk more than once, minimizing the slow disk-to-NPU transfers.  
For example, in a DP2TP2EP4 configuration, only one DP instance's attention weights are read from disk; the other DP instance relies on faster P2P transfers without additional disk I/O.

\begin{figure}[h]
  \centering
  \includegraphics[width=0.9\columnwidth]{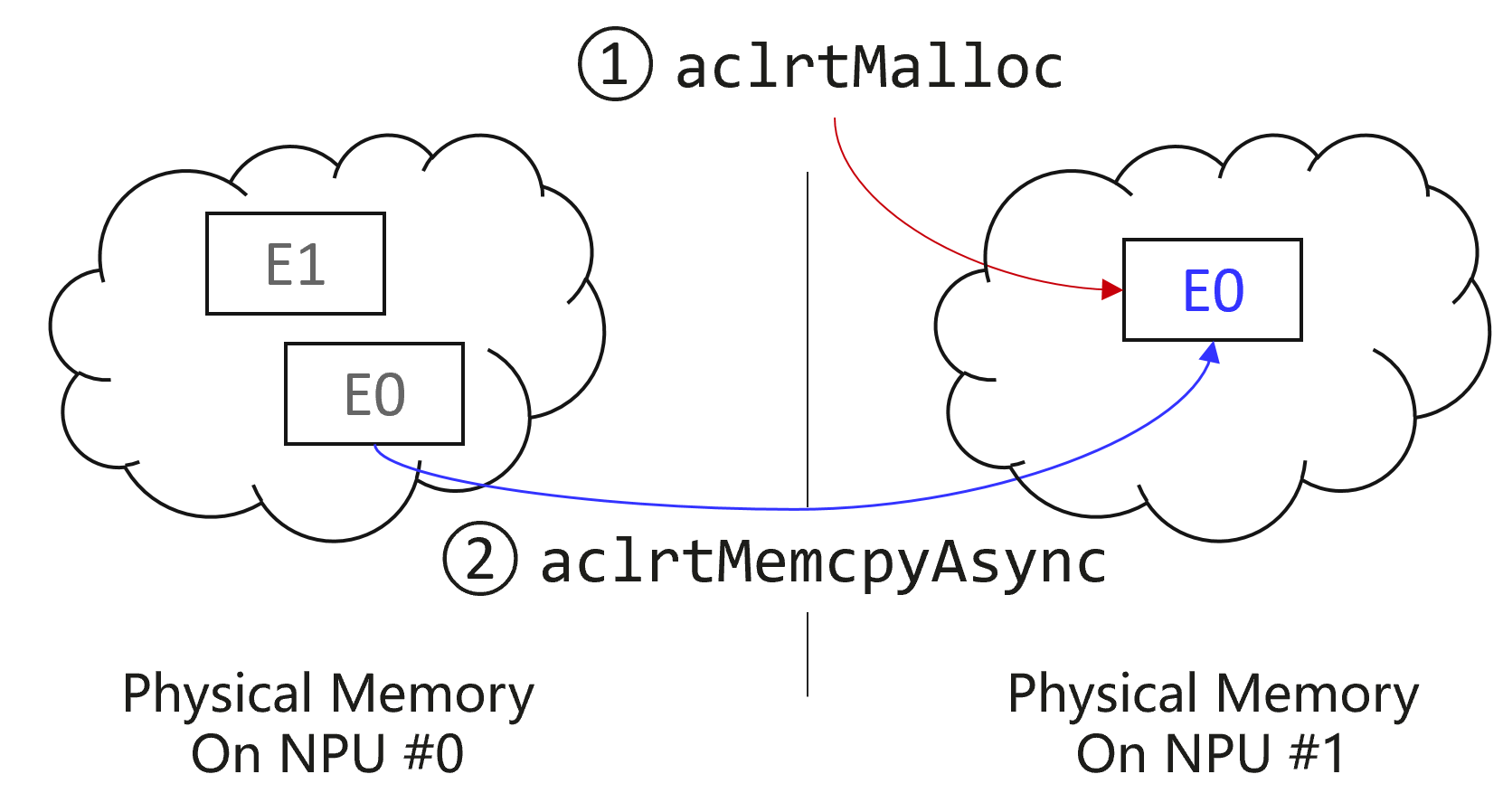}
  \caption{Peer-to-Peer (P2P) copy process. The target NPU allocates memory via \texttt{aclrtMalloc}, after which data is transferred asynchronously using \texttt{aclrtMemcpyAsync} across NPUs over the Ascend Unified Bus or equivalent interconnect.}
  \label{fig:p2p-copy}
\end{figure}

\begin{figure}[bp]
  \centering
  \includegraphics[width=0.9\columnwidth]{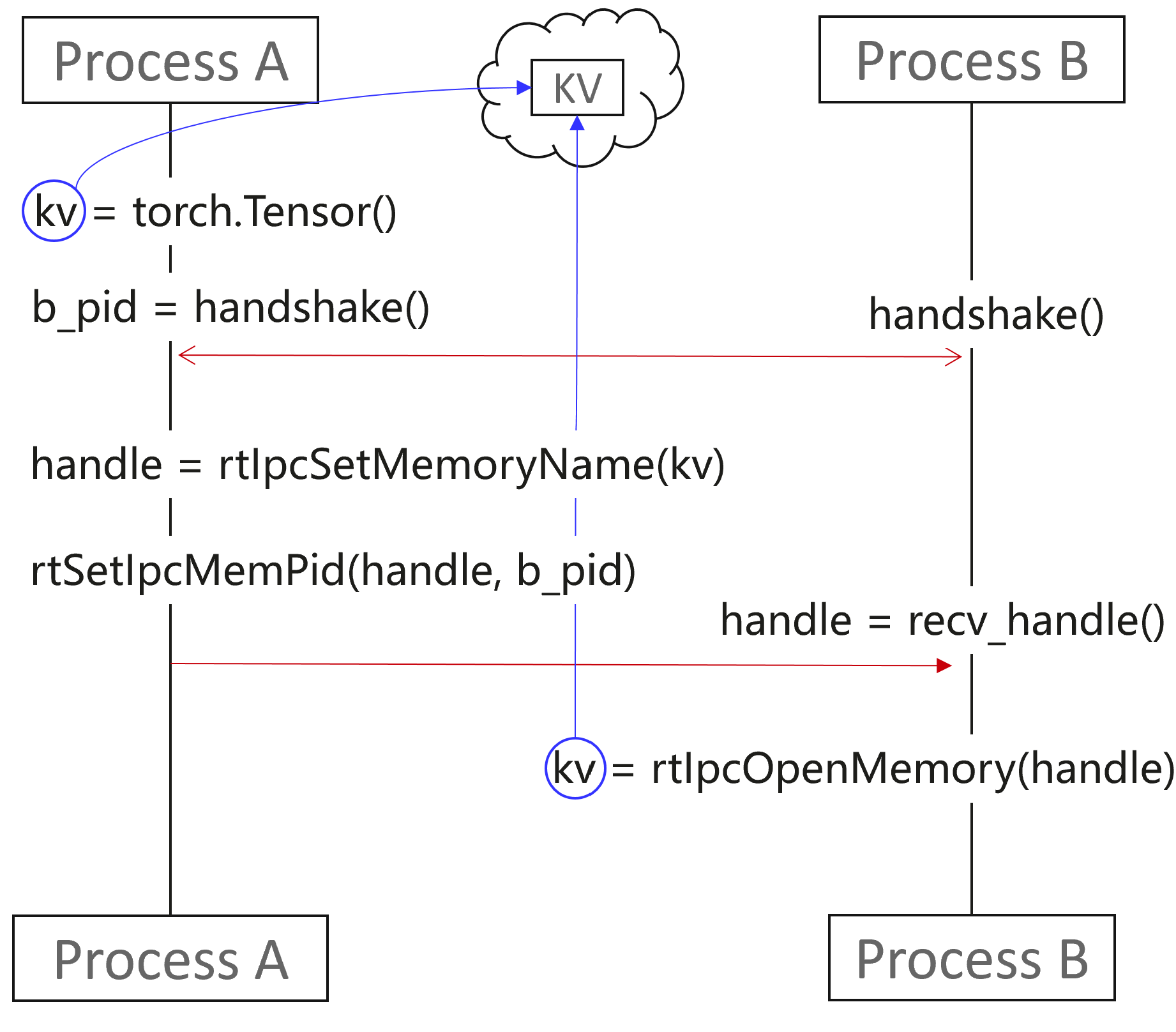}
  \caption{Zero-copy process. A tensor allocated in Process A is shared directly with Process B without duplication. The memory handle is registered via \texttt{rtIpcSetMemoryName} and shared with the destination process through IPC. Process B imports it with \texttt{rtIpcOpenMemory}, allowing both processes to reference the same physical memory.}
  \label{fig:zero-copy}
\end{figure}

\subsection{Fast Peer-to-Peer Tensor Transfer (\texttt{p2p-copy})}

To avoid costly disk I/O during scale-up, we define a high-speed peer-to-peer (P2P) transfers to move weights between NPUs. The \texttt{p2p-copy} primitive uses the Ascend Unified Bus or similar high-bandwidth interconnects to achieve this transfer efficiently. Specifically, a target NPU receives the required weights from a source NPU via an asynchronous transfer initiated by the \texttt{aclrtMemcpyAsync} API. This operation involves allocating the destination tensor on the target device and performing direct memory-to-memory transfer, bypassing host memory entirely. Optionally, it can be started on a separate stream to avoid blocking existing computation and memory operations inside the current NPU context. Because P2P transfers are typically an order of magnitude faster than disk I/O, this primitive is the preferred method for weight propagation during scale-out.

\subsection{Zero-Copy Sharing Across Processes (\texttt{zero-copy})}
To support sharing or reference-copy of a memory across independent processes, we implement a \texttt{zero-copy} primitive that allows a tensor allocated in a source process to be shared with a newly spawned destination process, effectively passing a reference without any memory duplication.
Speficailly, this is achieved by exporting the tensor memory handle via \texttt{rtIpcSetMemoryName()} and whitelisting the destination process using \texttt{rtSetIpcMemPid()}. The handle is transmitted through an IPC channel (e.g., a UNIX domain socket), and the receiver imports the tensor using \texttt{rtIpcOpenMemory()}. The physical pointer is then wrapped into a PyTorch tensor using \texttt{torch::from\_blob()}. 
Because this process avoids any actual data transfer, it is significantly faster than P2P copying and helps reduce peak memory pressure on shared NPUs. It is the core mechanism that enables concurrent scaling and inference without service interruption.

\begin{figure}[h]
  \centering
  \includegraphics[width=\columnwidth]{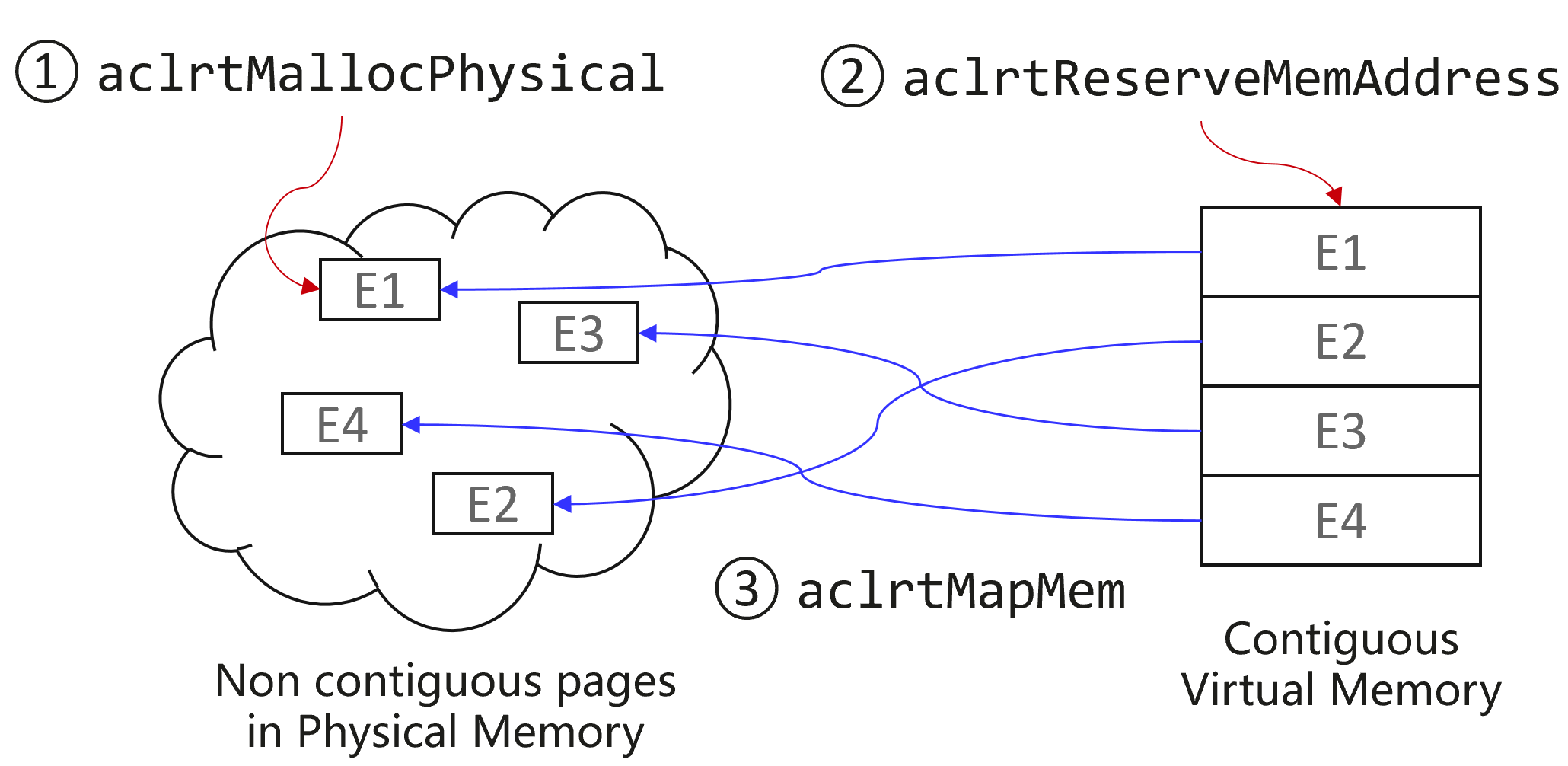}
  \caption{Virtual page allocation and remapping process. The primitive first allocates non-contiguous physical pages via \texttt{aclrtMallocPhysical}, reserves a contiguous virtual address range with \texttt{aclrtReserveMemAddress}, and then maps the physical pages into the virtual space using \texttt{aclrtMapMem}. This enables kernels to access expert weights as if they were contiguous in memory while preserving flexibility in the underlying physical placement.}
  \label{fig:virt-management}
\end{figure}

\subsection{Virtual Page Allocation and Remapping (\texttt{vpage-remap})}
To enable virtual memory–based expert weight management, ElasticMoE implements a primitive that allocates physical memory pages and binds them to a contiguous region in virtual address space. Using Ascend ACL memory APIs, the \texttt{vpage-remap} primitive first reserves the required contiguous virtual address range for all experts assigned to a device. It then allocates individual physical pages for each expert and binds them into the corresponding virtual offsets, allowing the logical layout to appear contiguous to kernels while the underlying physical placement remains flexible.

During scaling, when experts are migrated to or from a device, \texttt{vpage-remap} updates the virtual–physical mapping to point to the new pages—either locally allocated or received via \texttt{p2p-copy}—without reallocating or reshuffling the entire buffer. This remapping is performed asynchronously to allow the old inference instance to continue using the existing mappings until the new instance is fully activated. Once the transition completes, unused physical pages are unbound and released, minimizing peak memory usage.

\subsection{Adding new nodes to HMM (\texttt{add-nodes})}
Dynamically expands the number of nodes and NPUs managed by the HMM at runtime. A new node is first joined to the Ray cluster using its standard scaling API. The existing HCCL process group is then torn down via \texttt{destroy\_\allowbreak process\_\allowbreak group}, new Ray workers are launched for the added devices, and finally HCCL is reinitialized with \texttt{init\_\allowbreak process\_\allowbreak group} over the enlarged device set. This allows ElasticMoE to elastically grow cluster resources without restarting the system.

\section{Scale Down Operation} \label{scale-down-operation}

A scale-down operation in ElasticMoE reduces the number of NPUs allocated to an active inference instance, typically in response to declining load. For example, the system may transition from a configuration of DP2-TP2-EP4 across NPUs \texttt{0--3}, to DP1-TP2-EP2 on NPUs \texttt{0--1}.

Similar to scale-up, the Coordinator initiates the transition by signaling the HMM and IMM, which then execute their respective sub-tasks in parallel.

The HMM computes a weight reconfiguration plan that prioritizes zero-copy reuse and minimizes unnecessary P2P transfers. Since the tensor parallelism degree is fixed, both the attention weights and KV cache on NPUs \texttt{0--1} remain valid and reusable. No copying or reallocation is required for these components.

The only reconfiguration needed involves the expert layers. Expert weights residing on NPUs \texttt{2--3} must be transferred to NPUs \texttt{0--1}. To support this, new physical memory pages are allocated for incoming experts, followed by P2P transfers of the corresponding weights. Once copied, ElasticMoE remaps the virtual memory backing the expert blocks to the new physical pages using virtual memory primitives provided by ACL/CANN, maintaining the contiguous layout expected by inference kernels.

On the IMM side, the appropriate inference instance is retrieved or spawned, attaches to the newly mapped weights and cache via zero-copy, and signals readiness to the Coordinator. The Coordinator then transitions traffic to the scaled-down instance once all in-flight requests on the old instance have completed.

Overall, scale-down is a symmetric but simpler operation than scale-up, typically requiring fewer weight movements and incurring even lower peak memory overhead.

\end{document}